\documentclass[a4paper,fleqn]{cas-sc}
\let\printorcid\relax
\ExplSyntaxOn
\cs_set:Npn \__first_footerline:
  { \group_begin: \small \sffamily \__short_authors: \group_end: }
\ExplSyntaxOff

\usepackage{amssymb}
\usepackage{amsmath}
\usepackage{graphicx}
\usepackage{xcolor}
\usepackage[normalem]{ulem}
\newcounter{changecount}

\usepackage[numbers]{natbib}
\usepackage{algorithm}
\usepackage{algorithmic}

\begin{document}
\let\WriteBookmarks\relax
\def\floatpagepagefraction{1}
\def\textpagefraction{.001}
\hypersetup{hidelinks}

\shorttitle{Benchmarking quantum-annealing-based approaches for factory layout planning}

\shortauthors{Wu et al.}

\title[mode=title]{Benchmarking quantum-annealing-based approaches for factory layout planning}

\author[1]{Xiangqian Wu}
\cormark[1]
\ead{xiangqian.wu@rptu.de}

\author[1]{Philipp Schworm}
\author[1]{Marcel Wagner}
\author[1]{Jan C. Aurich}

\cortext[1]{Corresponding author}

\affiliation[1]{organization={Institute for Manufacturing Technology and Production Systems, RPTU Kaiserslautern},
                city={Kaiserslautern},
                country={Germany}}

\begin{abstract}
The early stage of factory layout planning has a direct effect on the operating costs, efficiency, and productivity of manufacturing systems. Conventional automated planning approaches often face a trade-off between solution quality and computational effort. Recent studies have investigated the potential of quantum annealing (QA) for solving factory layout planning problems. However, existing studies have mainly focused on individual QA-based formulations, whereas current hybrid QA solvers support multiple model types that combine QA and classical methods. To explore the potential of QA for factory layout planning, this paper benchmarks four QA-based approaches implemented with the corresponding solvers, including an iterative binary quadratic model (IBQM), a hybrid discrete quadratic model (HDQM), a hybrid constrained quadratic model (HCQM), and a hybrid nonlinear model (HNLM). To improve scalability, an iterative decomposition strategy is developed for IBQM, while a candidate selection procedure is introduced for HDQM. The approaches are benchmarked against  genetic algorithm, particle swarm optimization, and a hybrid tabu search and simulated annealing approach on $14$ layout instances with different layout sizes, geometric complexity, spatial restrictions, and transport route configurations. The comparative analysis considers solution quality, computation time, and scalability. The results show that QA-based approaches achieve competitive solution quality and substantial computational advantages, reducing computation time by up to $94.51\%$ relative to the classical reference methods. Among them, IBQM obtains the best solution quality in $78.6\%$ of the instances, with a maximum improvement of $17.41\%$ associated with an $81.34\%$ reduction in computation time. These findings provide a basis for their further application in industrial layout planning.
\end{abstract}


\begin{keywords}
Factory layout planning \sep Quantum annealing \sep Benchmarking \sep Combinatorial optimization
\end{keywords}

\maketitle


\section{Introduction}

Factory layout planning is a central part of factory planning. It defines the spatial arrangement of functional units within a manufacturing system, such as machines and support functions, with the objective of optimizing area utilization, production processes, and material flow \cite{wiendahl2023handbuch}.
The planning process is commonly divided into concept planning and detailed planning. In the early stage, concept planning creates a block layout that defines the relative positions and approximate sizes of functional units while considering constraints such as available space, building structure, and possible transport routes \cite{schenk2009factory}. This stage enables an initial assessment of the layout with objectives such as minimizing material handling cost and transport intensity, and it provides the foundation for detailed planning. In the detailed planning phase, the block layout is refined to the final factory layout by determining the exact positions of individual resources, buffers, and auxiliary equipment based on technical, organisational, and infrastructural details that are not considered in the conceptual layout \cite{wiendahl2023handbuch}.

The quality of layout generated in the early stage therefore has a direct influence on the operational performance and costs of manufacturing systems. Approaches for generating block layouts at this stage are generally classified into manual planning methods and automated planning methods. Manual planning is often supported by graphical construction procedures such as the Schmigalla triangle method, which arranges functional units iteratively on the basis of material flow or closeness relations \cite{schmigalla1970methoden}. Although these methods can produce feasible layout variants, they are time-consuming, explore a limited solution space, and depend on the planner's expertise \cite{banduka2017designing}. To address these limitations, automated planning approaches are used for systematic layout generation. These approaches formulate the layout problem as a combinatorial optimization problem known as the facility layout problem, which has been proven to be NP-hard \cite{sahni1976p}. With increasing problem size and complexity, exact methods quickly become impractical due to the exponential growth in computational effort. Metaheuristic methods are widely used to solve the layout planning problem \cite{drira2007facility}. However, for industrial-scale problems, conventional metaheuristic methods require substantial computation times to achieve high-quality solutions, while reducing computation time limits the exploration of the solution space and thereby reduces the solution quality \cite{rajwar2023exhaustive}.
Recent studies have indicated the potential of quantum annealing (QA) to address this limitation. Compared with classical metaheuristics, QA utilizes quantum-mechanical effects, including superposition, entanglement, and quantum tunneling, thereby enabling more efficient exploration of the solution space \cite{kadowaki1998quantum}. 
With advances in QA hardware, various QA-based formulations, including the Binary Quadratic Model (BQM), Constrained Quadratic Model (CQM), Discrete Quadratic Model (DQM), and Nonlinear Model (NLM), supported by corresponding hybrid quantum-classical solvers, have been developed. In other application areas, such as job shop scheduling, comparative studies have evaluated these formulations and demonstrated superior performance in solution quality and computation time compared with classical methods, especially for large problem instances \cite{schworm2023fjsp,schworm2024multiobjective,osaba2024nlhybrid,kim2025quantum,wu2026scheduling}. In factory layout planning, previous studies have shown the feasibility of QA-based approaches \cite{klar2022factory,wu2025factory}. In particular, the application of a CQM-based approach to an industrial factory use case showed that QA can be applied to practical layout problems \cite{wu2025factory}. Nevertheless, existing studies focus only on a single QA-based formulation and a specific factory layout problem. There is neither a comparison between different QA-based approaches nor a performance evaluation across manufacturing systems with varying characteristics, such as different layout sizes and geometries of functional units. Moreover, comparative studies between QA-based approaches and widely used metaheuristics under a common evaluation protocol remain limited, making it difficult to derive generalizable conclusions regarding the practical applicability of QA-based approaches to factory layout planning.

To address these gaps, this paper presents a systematic benchmarking study of QA-based approaches for factory layout planning. Different QA-based formulations are developed and implemented using their corresponding hybrid quantum-classical solvers. The resulting QA-based approaches are evaluated on factory layout instances with different characteristics and compared with widely used classical metaheuristics under a common evaluation protocol. This study provides a basis for assessing the suitability and practical applicability of QA-based methods for factory layout planning. The main contributions of this paper are summarized as follows:
\begin{enumerate}
    \item Four QA-based formulations for early-stage factory layout planning are developed, covering BQM, DQM, CQM, and NLM.
    \item Formulation-specific model refinements are introduced to improve the scalability and computational performance of QA-based approaches.
    \item A systematic benchmarking analysis is conducted by comparing the resulting QA-based approaches with widely used classical metaheuristics across factory layout instances with different characteristics.
\end{enumerate}
The remainder of this paper is organized as follows. Section~\ref{sec:sota} reviews automated factory layout planning approaches, introduces the principles of QA, and identifies the research gap. Section~\ref{sec:approach} presents the proposed QA-based factory layout planning framework, including the problem definition, the mathematical formulations of BQM, DQM, CQM, and NLM, and the corresponding model refinements. Section~\ref{sec:results} describes the numerical case study and discusses the comparative benchmarking results. Section~\ref{sec:conclusion} concludes the paper and outlines future work.
\section{State of the art}
\label{sec:sota}
\subsection{Automated factory layout planning approaches}
Despite the high computational complexity of factory layout planning, a wide range of automated approaches has been proposed to generate acceptable layout solutions within reasonable computation times \cite{perez2021facility}. These approaches can be classified into exact methods, which aim to identify globally optimal solutions, and approximate methods, which seek near-optimal solutions. Since combinatorial complexity increases exponentially with the number of functional units and layout positions, exact methods are limited to small-scale instances or simplified problem formulations \cite{bouramtane2024comprehensive}. For example, Palubeckis \cite{palubeckis2012branch} proposed a branch and bound algorithm for a simplified facility layout problem with fixed row locations, which solved an instance with 10 functional units in 1 s, whereas solving an instance with 35 functional units required 459 h, i.e., more than 19 days. Therefore, approximate approaches are commonly applied to large-scale problems. Approximate methods further include construction algorithms, improvement algorithms, and metaheuristic methods. More recently, artificial-intelligence-based approaches such as reinforcement learning have also been explored for layout planning \cite{klar2024transferable}. Early heuristic methods are mainly based on construction algorithms, which generate layout variants from scratch by sequentially placing functional units \cite{drira2007facility}. Although these methods are computationally simple, their solution quality is often limited because early placement decisions strongly constrain subsequent arrangements. To improve solution quality, improvement-based heuristic methods such as CRAFT \cite{banduka2017designing} start with an initial feasible solution and optimize it by iteratively changing the locations of the functional units \cite{nordin2016heuristics}. Although these methods are computationally efficient, their search processes are restricted by the local exchange mechanism. As a result, the solution quality of these methods is highly dependent on the initial layout and the selected neighborhood structure \cite{singh2006review}.
To enlarge the search space and improve the solution quality, metaheuristic methods are widely used to approximate the solution for industrial-scale layout problems. The most widely applied metaheuristics include simulated annealing (SA), genetic algorithms (GA), and particle swarm optimization (PSO) \cite{perez2021facility}. 
SA, inspired by the physical annealing process in metallurgy, probabilistically accepts worse solutions according to a cooling schedule, thereby enabling the search to escape local minima \cite{kirkpatrick1983optimization}. Chen et al. \cite{chen2025oscillation} proposed an SA-based algorithm for the single row facility layout problem, in which unequal-length functional units are arranged in one row along a straight line to minimize the transport intensity, showing applicability to large instances. However, the computational time increased substantially with problem size, rising from about 1s for instances with 110 feasible candidates to almost 40 min for instances with 2000 feasible candidates. GA improves a population of candidate solutions through selection, crossover, and mutation \cite{lambora2019genetic}. Hou et al. \cite{hou2019application} adopted GA to solve the unequal area facilities layout problem in a two-dimensional layout space. Although the approach was validated on benchmark problems up to 62 functional units, its population-based search still leads to increasing computational effort for larger instances. PSO, inspired by the collective behavior of bird flocks, iteratively adjusts candidate solutions based on individual and swarm-level information \cite{kennedy1995particle}. Liu et al.~\cite{liu2018multi} applied a multi-objective PSO algorithm to the unequal-area facility layout problem in a continuous flow shop manufacturing system, considering material handling cost, area utilization, and adjacency value for problems with up to 62 functional units. However, the attainment of high-quality solutions was associated with increasing computation times, thereby restricting the exploration of the solution space for more complex problem instances.
Accordingly, current classical metaheuristic methods still face the challenge of balancing the solution quality and computational time to solve the factory layout problems, especially with increasing size and complexity. As the number of functional units grows and additional constraints such as unequal areas, irregular geometries, orientation flexibility, minimum distances, and infrastructure limitations are considered, the solution space expands combinatorially. In industrial applications of factory layout planning, this trade-off between solution quality and computational effort directly influences the quality of the generated layouts and further affects the operational performance of the manufacturing system. Therefore, there remains a need for approaches that can produce high-quality layouts with low computational effort. In this context, QA may offer the potential to overcome this limitation.
\subsection{Quantum annealing}
\label{subsec-qa}
QA is one of the metaheuristic methods that exploits quantum mechanical effects to find near-optimal solutions for combinatorial optimization problems. The method formulates the optimization problem as the Hamiltonian energy minimization problem, in which each candidate solution corresponds to a physical state of the quantum system with an energy value. The aim is to find the ground state, i.e., the state with minimal energy, which corresponds to the optimal or near-optimal solution of the original problem \cite{kadowaki1998quantum}.
QA is physically realized through a controllable quantum system whose state is defined by quantum bits (qubits) influenced by external magnetic fields. At the beginning of the annealing process, the system is initialized in a superposition of all possible states by applying a strong transverse magnetic field. The system Hamiltonian is transformed from the initial configuration to the problem-specific Hamiltonian that encodes the energy state of the optimization problem, which is formulated in Eq.~\eqref{eq:Ham} with the binary variables $x_i, x_j$ and corresponding scalar weights $Q_{ii}, Q_{ij}$.

\begin{equation}
\label{eq:Ham}
H=\sum_{i}Q_{ii}x_i+\sum_{i<j}Q_{ij}x_ix_j
\end{equation}

As the first commercial company providing cloud access to QA hardware, D-Wave \footnote{Naming of specific companies is done solely for the sake of completeness and does not necessarily imply an endorsement of the named companies nor that the products are necessarily the best for the purpose.} currently provides QA systems with more than 5,000 qubits \cite{King.2023}. Since the problem must be mapped onto the physical qubit topology, the embedded problem size is limited. To address the scalability limitation, various hybrid solvers have been developed by combining classical and quantum resources. Table~\ref{tab:solver_comparison} summarizes the characteristics of the hybrid solvers as currently available \cite{dwave2023HybridQuadratic,dwaveNLPHybrid}. The D-Wave Hybrid Solver supports only BQM formulation and follows a decomposition-based solution process that combines tabu-search (TB), SA, and QA. Its maximum problem size is therefore not bounded by a fixed number of variables. In contrast, the Leap Hybrid Solvers support BQM, DQM, CQM, and NLM formulations and solve complete optimization problems using hybrid quantum-classical solution methods from D-Wave’s cloud platform \cite{dwaveLeapHybridDocs}.
\begin{table}[width=\linewidth,cols=6,pos=t]
\caption{Comparison of different hybrid solvers.}
\label{tab:solver_comparison}
\small
\begin{tabular*}{\tblwidth}{@{}LLLLLL@{}}
\toprule
 & D-Wave Hybrid Solver & \multicolumn{4}{c}{Leap Hybrid Solvers} \\
\cmidrule(lr){2-2} \cmidrule(l){3-6}
Model & BQM & BQM & DQM & CQM & NLM \\
\midrule
Variable types & Binary & Binary & Discrete & Binary, integer, real & Binary, integer \\
Max. variables & -- & 1,000,000 & 5,000 & 5,000,000 & -- \\
Max. cases & -- & -- & 500,000 & -- & -- \\
Objective terms & Quadratic & Quadratic & Quadratic & Quadratic & Nonlinear \\
Constraint handling & Penalty terms & Penalty terms & Penalty terms & Explicit & Explicit \\
Max. constraints & -- & -- & -- & 100,000 & -- \\
Max. nodes & -- & -- & -- & -- & 2,000,000 \\
Max. decision state size & -- & -- & -- & -- & 52,428,800 \\
\bottomrule
\end{tabular*}
\end{table}
BQM and DQM are quadratic formulations in which constraints are incorporated into the objective function through weighted penalty terms. BQM uses only binary variables, whereas DQM supports further discrete variables in which one discrete variable can represent multiple cases. The Leap Hybrid BQM and DQM solvers support up to 1,000,000 variables and 5,000 variables with 500,000 total cases, respectively. Compared to the unconstrained quadratic formulations, CQM and NLM implement the constraints individually, reducing the need to adjust penalty weights. The Leap Hybrid CQM solver supports binary, integer, and continuous variables with up to 5,000,000 variables and 100,000 constraints. NLM extends CQM by supporting nonlinear terms in both the objective and constraints, including higher-order polynomial expressions and specialized decision variable types. Accordingly, the scalability of the Leap Hybrid nonlinear solver is characterized by the node limit and decision-state-size limit, with up to 2,000,000 nodes and a maximum decision state size of 52,428,800 bytes.
 
The availability of these solvers with different modeling capabilities has motivated comparative studies evaluating their performance characteristics. Osaba et al. \cite{osaba2024nlhybrid} benchmarked the NLM solver against the BQM and CQM solvers on 45 instances across three combinatorial optimization problems, including the Traveling Salesman Problem, the Knapsack Problem, and the Maximum Cut Problem. The results demonstrated that the NLM solver achieved statistically superior results overall, while the BQM solver performed better on problems dominated by binary decision variables. Kim et al.~\cite{kim2025quantum} proposed a benchmarking study on material optimization and random dense quadratic unconstrained binary optimization problems, solving D-Wave Hybrid Solver and classical methods, including SA, TB, steepest descent, and integer programming. The results showed the advantage of QA over classical methods for achieving high solution quality and reduced computation time, especially in large-scale optimization problems. 
Quinton et al. \cite{quinton2025quantum} conducted a comprehensive evaluation of the Leap Hybrid CQM solver against industry-leading classical solvers across diverse optimization problem categories and found that the performance of QA-based approaches depends strongly on problem structure and formulation quality.
In the context of industrial applications, QA-based approaches have been explored for diverse application areas. Schworm et al.~\cite{schworm2023fjsp} proposed various QA-based formulations to solve flexible job shop scheduling problems across different instance sizes and evaluated several hybrid QA solvers, showing that the iterative QA approach based on the D-Wave Hybrid solver outperformed the other QA solvers and the classical benchmark methods. Its application to multi-objective optimization and industrial use cases further indicated the practical potential of QA in production-related decision problems \cite{schworm2024multiobjective,SCHWORM2026100132}. Sakuler et al.~\cite{sakuler2025portfolio} applied QA to portfolio optimization using the Leap Hybrid CQM and Leap Hybrid BQM solvers. The comparison with classical methods showed that both QA-based approaches achieved higher solution quality, while the Leap Hybrid CQM solver outperformed the Leap Hybrid BQM solver for the considered problem. These studies show that QA-based performance depends strongly on the selected formulation, solver architecture, and problem characteristics.

\subsection{Research gap}

Existing studies have demonstrated the feasibility of applying QA to factory layout planning and have reported promising potential for industrial applications \cite{klar2022factory,wu2025factory}. However, current research largely focuses on the first applications and development of single QA-based approaches, while systematic comparisons across various QA methods and comparative studies with traditional optimization methods remain limited. The study in \cite{klar2022factory} proposed the first QA-based factory layout planning using a BQM formulation with binary variables to represent the assignment of functional units to predefined candidate positions. The approach demonstrated the feasibility of mapping factory layout problems onto a quantum annealer and showed promising scalability characteristics. However, the study simplified the layout problem and did not compare with classical optimization methods. Wu et al.~\cite{wu2025factory} extended this work by introducing a CQM formulation with continuous and binary variables that enables continuous placement and flexible orientation decisions in factory layout planning and offers improved scalability. The CQM-based approach was applied to an industrial use case and compared with the non-dominated sorting genetic algorithm II (NSGA-II), demonstrating superior performance in terms of both transport intensity minimization and area utilization maximization. However, the evaluation of QA performance was limited to a single industrial scenario, and only one QA-based formulation was considered. Since the performance of QA-based approaches can vary with formulation, solver configuration, problem size, and instance characteristics, a scalable and comparable assessment is required to evaluate industrial applicability. Consequently, this paper addresses this gap by benchmarking various QA-based approaches against widely used classical metaheuristic methods on use cases of different sizes and characteristics, using a common protocol to evaluate performance in terms of solution quality and computation time.
\section{Factory layout planning using quantum annealing}
\label{sec:approach}
\subsection{Framework}
\begin{figure}[pos=t]
\centering
\includegraphics[width=.9\linewidth]{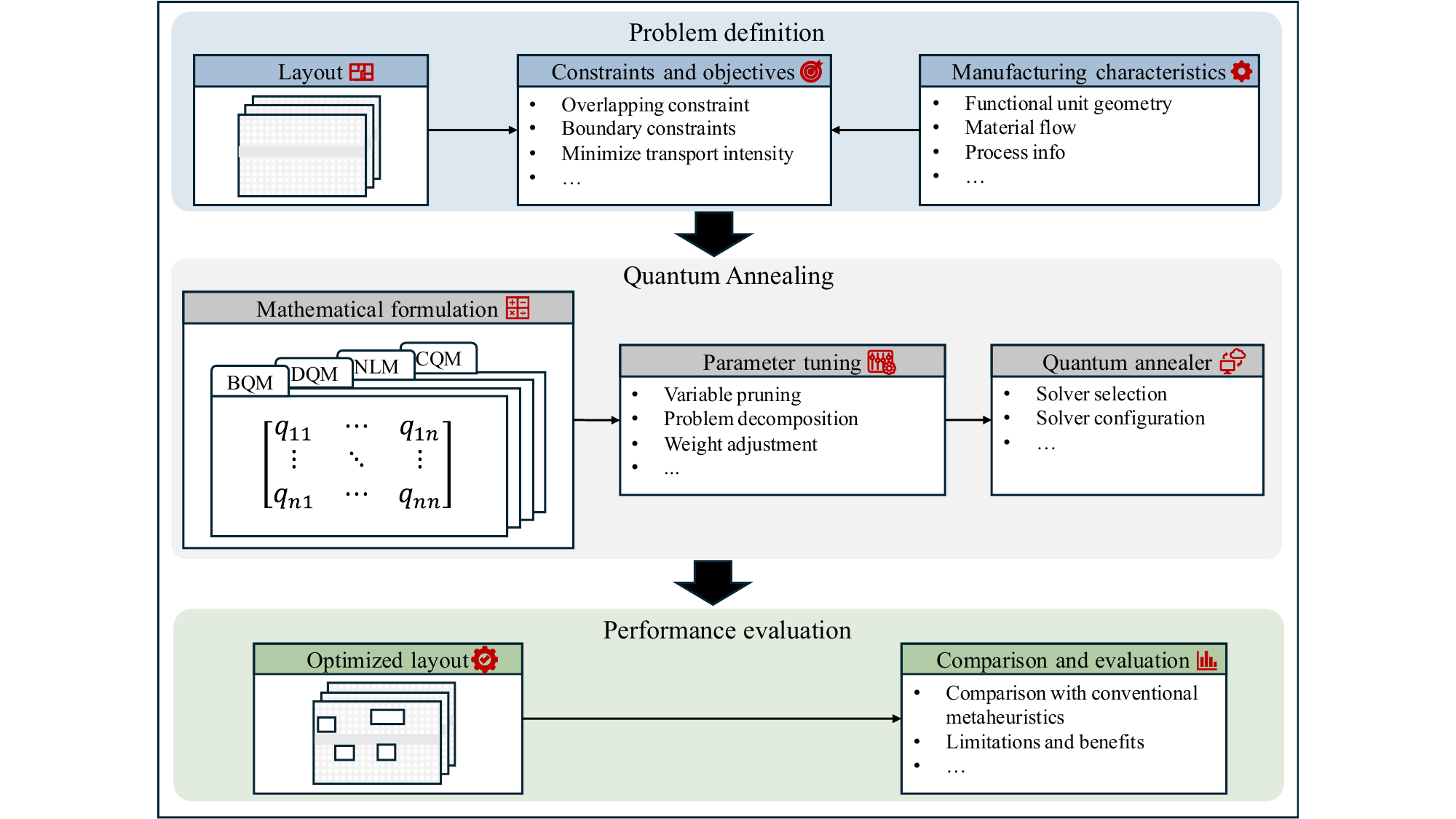}
\caption{Framework for QA-based factory layout planning.}
\label{fig:framework}
\end{figure}

Fig.~\ref{fig:framework} presents the framework for benchmarking QA-based factory layout planning approaches, including the modeling workflow and the performance evaluation. The framework is designed to evaluate the scalability and applicability of different QA-based formulations and solver configurations under the same problem definition and a consistent evaluation protocol. 
In the initial step, a layout instance is defined by the floor area size, the transport routes, and the infeasible placement areas, while the manufacturing characteristics are specified by the functional units and their interrelationships, represented by a material flow matrix and a distance matrix. Based on these inputs, the optimization problem is defined through constraints and objectives. The constraints describe feasibility requirements such as non-overlapping and minimum distance constraints, whereas objectives quantify layout quality, for example, through the minimization of transport intensity. After the problem definition, the layout problem is formulated into QA-based mathematical formulations. Since different QA-based solvers have different modeling requirements, this framework considers BQM, DQM, CQM, and NLM formulations. Each model is further refined to improve scalability and solution quality. The refinements include variable pruning through the choice of variable types and decomposition strategies to improve scalability, as well as weight adjustments to ensure constraint satisfaction and optimization quality. Afterward, the refined models are solved using the corresponding QA solvers, resulting in the QA-based approaches denoted as IBQM for the BQM formulation solved with the D-Wave Hybrid Solver, and HDQM, HCQM, and HNLM for the DQM, CQM, and NLM formulations solved with the corresponding Leap Hybrid solvers, respectively.
In the final step, the resulting layouts are evaluated and compared with those obtained by conventional metaheuristics across benchmark instances with different sizes and characteristics. The evaluation focuses on solution quality, computational effort, scalability, and the limitations of each QA-based approach.
\subsection{Problem definition}
\begin{figure}[pos=t]
\centering
\includegraphics[width=.9\linewidth]{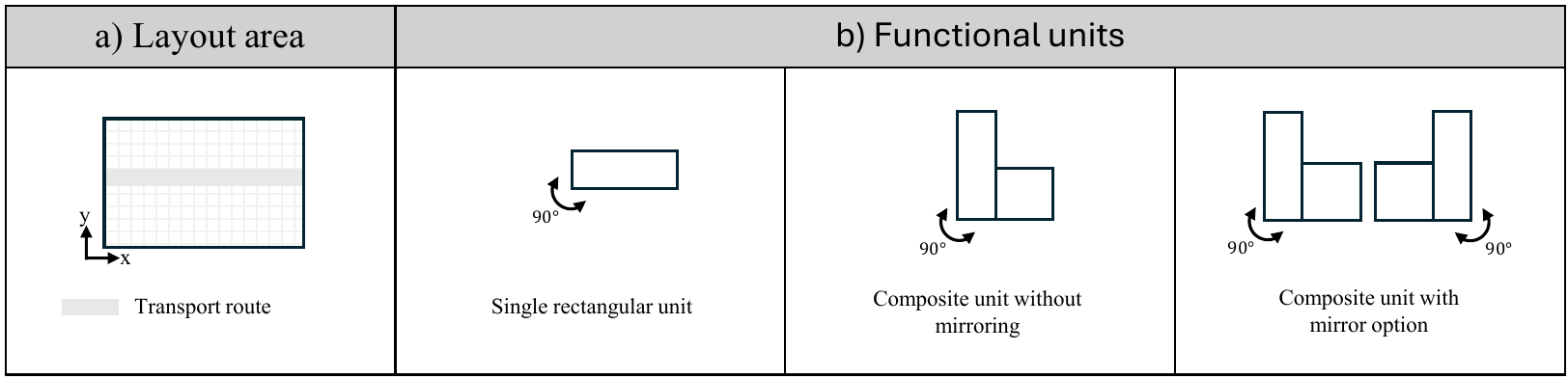}
\caption{Geometry of layout and functional units.}
\label{fig:Geolayout}
\end{figure}

The factory layout planning problem considered in this study is based on prior work \cite{wu2025factory}. In the early layout planning phase, the layout area and the functional units are represented by rectangles or combinations of rectangular geometries, as illustrated in Fig.~\ref{fig:Geolayout}.
The layout area is described in a two-dimensional coordinate system with the lower left corner as the reference point, where $X=\{0,\ldots,X_{\max}\}$ and $Y=\{0,\ldots,Y_{\max}\}$ denote the sets of all coordinates in the horizontal and vertical directions, respectively. Within this area, a set of functional units $U=\{1,\dots,n\}$ must be arranged. According to their geometric representation, the functional units are classified into single rectangular units, composite units without mirroring, and composite units with a mirror option. Each functional unit $u\in U$ consists of one or two rectangles, represented by the rectangle set $R_u$, where $|R_u|\in \{1,2\}$. The dimensions of each rectangle $i\in R_u$ are given by its length $l_{u,i}$ and width $w_{u,i}$. Single rectangular units with $|R_u|=1$ have two orientations, namely horizontal and vertical. For composite units with $|R_u|=2$, the predefined relative arrangement of the two rectangles is characterized by $\Delta x_{u,ij}$ and $\Delta y_{u,ij}$ for $i,j \in R_u$, together with the parameter $\mu_u$, where $\mu_u \in \{0,1\}$ indicates whether a mirror option is allowed. Furthermore, composite units can rotate in four directions, which are $0^\circ$, $90^\circ$, $180^\circ$, and $270^\circ$. For composite units with $\mu_u=1$, both the original arrangement and its mirrored arrangement are considered.
Moreover, the material flow from functional unit $u$ to $u'$ with $u,u'\in U$ is denoted by $M_{u,u'}$. The transport distance between functional units is defined by the Manhattan metric based on the center coordinates of the corresponding units. In addition to transport between functional units, the material flows from the layout input $I$ to functional unit $u$ and from functional unit $u$ to layout output $O$ are also considered, which are denoted by $M_{I,u}$ and $M_{u,O}$.
Considering the geometric characteristics of the functional units and the manufacturing scenario, the following three constraints are considered:
\begin{enumerate}
    \item \textbf{Non-overlapping and minimum-distance constraint:} Each functional unit must be placed without overlapping any other functional unit. Moreover, predefined minimum distances, denoted by $Dx_{u,u'}$ and $Dy_{u,u'}$, in the horizontal or vertical direction between two functional units with $u,u'\in U$ have to be satisfied for manufacturing safety and material transport.
    \item \textbf{Area constraint:} All functional units must be placed within the admissible layout area and must not overlap with the transport route area $TP$ or the infeasible placement area $IM$.
    \item \textbf{Combination constraint:} For a composite unit $u$ with $|R_u|=2$, the relative positions of two rectangles $i,j\in R_u$ must satisfy the predefined arrangement.
\end{enumerate}
            
In this study, the objective is to minimize transport intensity, including the transport between functional units as well as the transport between functional units and the layout input and output.
\subsection{Mathematical formulation}
To compare different QA-based approaches under the same problem definition, the factory layout planning problem is formulated as BQM, DQM, CQM, and NLM. 
\subsubsection{BQM formulation}
Since BQM supports only binary variables, the placement of each rectangle $i\in R_u$ from functional unit $u \in U$ is represented by the binary variable $v_{u,i,p}$ in Eq.~\eqref{eq:bqm_var}, where $p \in P_{u,i}$ denotes a placement candidate defined by the center coordinate $(x,y)$ and the orientation $k$. The set of feasible placement candidates $P_{u,i}$, excluding candidates that overlap with the infeasible placement area $IM$ or the transport route area $TP$, is given in Eq.~\eqref{eq:candi_ps}.
\begin{equation}
\label{eq:bqm_var}
v_{u,i,p}=
\left\{
\begin{array}{@{}l@{}l@{}}
1,\; \text{if rectangle } i\in R_u \text{ is assigned to } p \in P_{u,i}\\
0,\; \text{otherwise}
\end{array}
\right.
\end{equation}
with
\begin{equation}
\label{eq:candi_ps}
\begin{aligned}
P_{u,i}=\{\,p \mid\;& p=(x,y,k),\; x\in X,\; y\in Y,\; k\in\{0,1\},\; p\cap TP=\varnothing,\; p\cap IM=\varnothing\,\}
\end{aligned}
\end{equation}
Here, $k=0$ represents the original orientation and $k=1$ denotes the rotated orientation. Based on these binary variables, other problem-specific constraints and the objective are formulated as binary polynomials in the Hamiltonian. The non-overlapping and minimum-distance constraint is formulated by $h_1$ in Eq.~\eqref{eq:bqm_c1}, which penalizes all cases in which two rectangles from different functional units do not satisfy the required minimum distances in both coordinate directions, including overlapping cases. These infeasible cases are defined by the set $\mathcal{C}_{\mathrm{ov}}$ in Eq.~\eqref{eq:bqm_conflict}. 

\begin{equation}
\label{eq:bqm_c1}
h_1=\sum_{((u,i,p),(u',j,p'))\in \mathcal{C}_{\mathrm{ov}}} v_{u,i,p}\,v_{u',j,p'}
\end{equation}
with
\begin{equation}
\label{eq:bqm_conflict}
\mathcal{C}_{\mathrm{ov}}=
\left\{
\big((u,i,p),(u',j,p')\big)\;\middle|\;
\begin{aligned}
&u,u'\in U,\; u\neq u',\\
& i\in R_u,\; j\in R_{u'},\\
&p=(x,y,k)\in P_{u,i},\\
&p'=(x',y',k')\in P_{u',j},\\
&|x-x'|-\dfrac{l_{u,i}^{k}}{2}-\dfrac{l_{u',j}^{k'}}{2}<Dx_{u,u'},\\
&|y-y'|-\dfrac{w_{u,i}^{k}}{2}-\dfrac{w_{u',j}^{k'}}{2}<Dy_{u,u'}
\end{aligned}
\right\}
\raisetag{1.2ex}
\end{equation}
Here, $l_{u,i}^{k}$ and $w_{u,i}^{k}$ denote the length and width of rectangle $i \in R_u$ under orientation $k$. When $h_1=0$, the non-overlapping and minimum-distance constraint is satisfied. For a composite functional unit with $|R_u|=2$, the combination constraint is represented by $h_2$ in Eq.~\eqref{eq:bqm_c2}. Since the number of cases satisfying the required relative arrangement is substantially smaller than the number of infeasible cases, $h_2$ is formulated as a reward term. Accordingly, negative quadratic coefficients are assigned to those cases in which the two rectangles from the same functional unit satisfy the predefined relative arrangement. The corresponding set is denoted by $\mathcal{C}^{\mathrm{comb}}_{u,ij}$, which is classified into the non-mirrored condition $\mathcal{C}^{\mathrm{nm}}_{u,ij}$ and the mirrored condition $\mathcal{C}^{\mathrm{m}}_{u,ij}$ according to $\mu_u$, as given in Eq.~\eqref{eq:bqm_comb}. In both cases, the two rectangles must share the same orientation. For $\mu_u=0$, their relative placement must satisfy the predefined arrangement under four rotational states of the composite unit (Eq.~\eqref{eq:bqm_comb_nm}). For $\mu_u=1$, the absolute relative placement must satisfy the predefined arrangement under the corresponding rotated orientation (Eq.~\eqref{eq:bqm_comb_m}).
\begin{equation}
\label{eq:bqm_c2}
h_2=
-\sum_{\substack{u\in U\\ |R_u|=2}}
\sum_{\substack{i,j\in R_u\\ i<j}}
\sum_{(p,p')\in \mathcal{C}^{\mathrm{comb}}_{u,ij}}
v_{u,i,p}\,v_{u,j,p'}
\end{equation}
\begin{equation}
\label{eq:bqm_comb}
\mathcal{C}^{\mathrm{comb}}_{u,ij}=
\begin{cases}
\mathcal{C}^{\mathrm{nm}}_{u,ij}, & \text{if } \mu_u=0,\\[4pt]
\mathcal{C}^{\mathrm{m}}_{u,ij}, & \text{if } \mu_u=1,
\end{cases}
\end{equation}
\begin{equation}
\label{eq:bqm_comb_rot}
(\Delta x^{k}_{u,ij},\Delta y^{k}_{u,ij})=
\begin{cases}
(\Delta x_{u,ij},\Delta y_{u,ij}), & k=0,\\
(-\Delta y_{u,ij},\Delta x_{u,ij}), & k=1.
\end{cases}
\end{equation}

\begin{equation}
\label{eq:bqm_comb_m}
\mathcal{C}^{\mathrm{m}}_{u,ij}=
\left\{
(p,p')
\;\middle|\;
\begin{aligned}
&p=(x,y,k)\in P_{u,i},\;
p'=(x',y',k')\in P_{u,j},\;
k'=k,\\
&(|x-x'|,|y-y'|)=
(|\Delta x^{k}_{u,ij}|,|\Delta y^{k}_{u,ij}|)
\end{aligned}
\right\}
\end{equation}
\begin{equation}
\label{eq:bqm_comb_nm}
\mathcal{C}^{\mathrm{nm}}_{u,ij}=
\left\{
(p,p')
\;\middle|\;
\begin{aligned}
&p=(x,y,k)\in P_{u,i},\;
p'=(x',y',k')\in P_{u,j},\;
k'=k,\\
&(x-x',y-y')=\pm (\Delta x^{k}_{u,ij},\Delta y^{k}_{u,ij})
\end{aligned}
\right\}
\end{equation}
Since the placement of each rectangle is described by a set of binary variables over its candidate placements, an additional assignment constraint is required to ensure that each rectangle is assigned to exactly one placement candidate. This constraint is expressed by $h_3$ in Eq.~\eqref{eq:bqm_c3}. It is satisfied when $h_3=0$.
\begin{equation}
\label{eq:bqm_c3}
h_3=\sum_{u\in U}\sum_{i\in R_u}\left(1-\sum_{p\in P_{u,i}} v_{u,i,p}\right)^2
\end{equation}
The objective of transport intensity minimization is denoted by $h_4$. The transport intensity is determined by the corresponding material flow and Manhattan distance between the related transport centers. As shown in Eq.\eqref{eq:bqm_o1}, $h_4$ includes three considered transport cases, namely transport between functional units, transport from the layout input to functional units, and transport from functional units to the layout output. The corresponding transport terms are defined in Eq.\eqref{eq:bqm_transport}.
\begin{equation}
\label{eq:bqm_o1}
\begin{aligned}
h_4={}&
\sum_{\substack{u,u'\in U\\u\neq u'}}
\sum_{\substack{i\in R_u\\j\in R_{u'}}}
\sum_{\substack{p\in P_{u,i}\\p'\in P_{u',j}}}
\tilde{T}_{u,u',i,j,p,p'}\,v_{u,i,p}v_{u',j,p'}\\
&+\sum_{u\in U}\sum_{i\in R_u}\sum_{p\in P_{u,i}}
\tilde{T}_{I,u,i,p}\,v_{u,i,p}
+\sum_{u\in U}\sum_{i\in R_u}\sum_{p\in P_{u,i}}
\tilde{T}_{u,O,i,p}\,v_{u,i,p}
\end{aligned}
\end{equation}
with
\begin{equation}
\label{eq:bqm_transport}
\begin{aligned}
&T_{u,u',i,j,p,p'}=M_{u,u'}\bigl(|x-x'|+|y-y'|)\\
&T_{I,u,i,p}=M_{I,u}\bigl(|x-x_I|+|y-y_I|)\\
&T_{u,O,i,p}=M_{u,O}\bigl(|x-x_O|+|y-y_O|)
\end{aligned}
\end{equation}
To balance the objective and the different constraint terms in the Hamiltonian, each transport term is normalized as shown in Eq.~\eqref{eq:bqm_norm}, resulting in the normalized transport term $\tilde{T}$, where $T_{\min}$ and $T_{\max}$ are the minimum and maximum transport values.
\begin{equation}
\label{eq:bqm_norm}
\tilde{T}=
\dfrac{T-T_{\min}}{T_{\max}-T_{\min}}
\end{equation}
The final BQM Hamiltonian, represented by $H_{BQM}$, is given in Eq.\eqref{eq:bqm_hamiltonian}, where the constraints and objective are combined with the corresponding non-negative weights $\omega_1, \omega_2, \omega_3, \omega_4$.
\begin{equation}
\label{eq:bqm_hamiltonian}
H_{\mathrm{BQM}}=\omega_1h_1+\omega_2h_2+\omega_3h_3+\omega_4h_4
\end{equation}
\subsubsection{DQM formulation}
Different from the BQM formulation, the placement of each functional unit $u\in U$ is represented by one discrete variable $z_u$, taking values in the case set $S_u$, as given in Eq.~\eqref{eq:dqm_var}. For a single rectangular functional unit with $|R_u|=1$, each case corresponds to one feasible placement candidate $p \in P_{u,i}$. For a composite functional unit with $|R_u|=2$, the case set consists of feasible placement pairs of the two rectangles satisfying the predefined relative arrangement, as given in Eq.~\eqref{eq:dqm_state}. 
\begin{equation}
\label{eq:dqm_var}
z_u \in S_u, \qquad u\in U
\end{equation}
with
\begin{equation}
\label{eq:dqm_state}
S_u=
\begin{cases}
P_{u,i}, & \text{if } |R_u|=1\\[6pt]
\left\{
(p,p')
\;\middle|
(p,p')\in \mathcal{C}^{\mathrm{comb}}_{u,ij}
\right\}, & \text{if } |R_u|=2,\; i,j\in R_u 
\end{cases}
\end{equation}

Similar to the BQM formulation, placement candidates overlapping with the infeasible placement area or the transport route area are excluded from the case set, which supports the satisfaction of the area constraint. Accordingly, the assignment, combination, and area constraints are satisfied by the definition of cases for the discrete variable. The case indicator $\chi_{u,s}$ in Eq.~\eqref{eq:dqm_case_indicator} specifies case $s$ of functional unit $u$.
\begin{equation}
\label{eq:dqm_case_indicator}
\chi_{u,s}(z_u)=
\begin{cases}
1, & \text{if } z_u=s\\
0, & \text{otherwise}
\end{cases}
\end{equation}
For each $i\in R_u$ and $s\in S_u$, the center coordinates and orientation of rectangle $i$ under case $s$ are denoted as $x_{u,i}(s)$, $y_{u,i}(s)$, and $k_{u,i}(s)$. The non-overlapping and minimum-distance constraint is represented by $h_1$, which directly penalizes case combinations of two different functional units whose corresponding placements overlap or do not satisfy the required minimum distances. The corresponding set $\Omega_{\mathrm{ov}}$ is defined in Eq.~\eqref{eq:dqm_conflict}.
\begin{equation}
\label{eq:dqm_c1}
h_1=
\sum_{((u,s),(u',s'))\in \Omega_{\mathrm{ov}}}
\chi_{u,s}(z_u)\chi_{u',s'}(z_{u'})
\end{equation}
with
\begin{equation}
\label{eq:dqm_conflict}
\Omega_{\mathrm{ov}}=
\left\{
\big((u,s),(u',s')\big)
\;\middle|\;
\begin{aligned}
&u,u'\in U,\; u\neq u',\; s\in S_u,\; s'\in S_{u'},\\
&\exists\, i\in R_u,\; j\in R_{u'},\\
&|x_{u,i}(s)-x_{u',j}(s')|
-\dfrac{l_{u,i}^{k_{u,i}(s)}}{2}
-\dfrac{l_{u',j}^{k_{u',j}(s')}}{2}
<Dx_{u,u'},\\
&|y_{u,i}(s)-y_{u',j}(s')|
-\dfrac{w_{u,i}^{k_{u,i}(s)}}{2}
-\dfrac{w_{u',j}^{k_{u',j}(s')}}{2}
<Dy_{u,u'}
\end{aligned}
\right\}
\raisetag{1.0ex}
\end{equation}

The objective of transport intensity minimization is denoted by $h_2$. Analogous to the BQM formulation, the corresponding transport terms are determined by the material flow and the Manhattan distance between the related transport centers and are also normalized according to Eq.~\eqref{eq:bqm_norm}. Based on the normalized transport terms, the objective term is formulated in Eq.~\eqref{eq:dqm_o1}. 
\begin{equation}
\label{eq:dqm_o1}
\begin{aligned}
h_2={}&
\sum_{\substack{u,u'\in U\\u\neq u'}}
\sum_{\substack{i\in R_u\\j\in R_{u'}}}
\sum_{\substack{s\in S_u\\s'\in S_{u'}}}
\tilde{T}_{u,u',i,j,s,s'}\,\chi_{u,s}(z_u)\chi_{u',s'}(z_{u'})\\
&+\sum_{u\in U}\sum_{i\in R_u}\sum_{s\in S_u}
\tilde{T}_{I,u,i,s}\,\chi_{u,s}(z_u)
+\sum_{u\in U}\sum_{i\in R_u}\sum_{s\in S_u}
\tilde{T}_{u,O,i,s}\,\chi_{u,s}(z_u)
\end{aligned}
\end{equation}
The final DQM Hamiltonian, represented by $H_{\mathrm{DQM}}$, is given in Eq.~\eqref{eq:dqm_h}.
\begin{equation}
\label{eq:dqm_h}
H_{\mathrm{DQM}}=\omega_1h_1+\omega_2h_2
\end{equation}
\subsubsection{CQM formulation}

The CQM formulation is based on prior work \cite{wu2025factory}. To support a
consistent evaluation of scalability across different solvers, the CQM formulation in this study introduces several modifications, thereby reducing the number of variables and constraints,
restricting the search space, and potentially improving solution quality,
particularly for large problem instances.
\begin{enumerate}
\item Integer position variables replace real-valued variables,
discretizing the search space to a unit coordinate grid.
\item Binary variables with domain $\{0,1\}$ are used for rotation and indicator variables.
\item Auxiliary transport variables are created only for pairs with
nonzero material flow, and auxiliary variables for non-overlapping and
minimum-distance constraints are reduced, thereby decreasing the number of
variables and linear terms.
\end{enumerate}
For each rectangle $i\in R_u$ of functional unit $u\in U$, the center
position and orientation are represented by the integer variables
$a_{u,i}$, $b_{u,i}$, and the binary variable $c_{u,i}$, as defined in
Eq.~\eqref{eq:cqm_var}.
\begin{equation}
\label{eq:cqm_var}
\begin{aligned}
a_{u,i} &\in \{0,1,\ldots,X_{\max}\}\\
b_{u,i} &\in \{0,1,\ldots,Y_{\max}\}\\
c_{u,i} &= \begin{cases}
1,&\text{if rectangle } i \in R_u \text{ is rotated by }90^\circ\\
0,&\text{if rectangle } i \in R_u \text{ is not rotated}
\end{cases}
\end{aligned}
\end{equation}
For the non-overlapping and minimum-distance constraints, prior work
Wu et al.~\cite{wu2025factory} introduced three binary auxiliary variables in each
coordinate for each rectangle pair. In contrast, this work uses four
inequalities (Eqs.~\eqref{eq:c1_1}--\eqref{eq:c1_4}) with two binary
auxiliary variables $\alpha_{u,i,u',j}$ and $\beta_{u,i,u',j}$ for each
rectangle pair $i\in R_u$, $j\in R_{u'}$, $u\neq u'$. These inequalities describe the four possible relative directions, of which at least one
must satisfy the required minimum separation, denoted by
$d^{x,\mathrm{min}}_{u,i,u',j}$ and $d^{y,\mathrm{min}}_{u,i,u',j}$.
\begin{align}
\label{eq:c1_1}
a_{u,i} - a_{u',j}
  + \alpha_{u,i,u',j}X_{\max} + \beta_{u,i,u',j}X_{\max}
&\geq d^{x,\mathrm{min}}_{u,i,u',j}\\
\label{eq:c1_2}
a_{u',j} - a_{u,i}
  + \alpha_{u,i,u',j}X_{\max} + (1-\beta_{u,i,u',j})X_{\max}
&\geq d^{x,\mathrm{min}}_{u,i,u',j}
\end{align}
\begin{align}
\label{eq:c1_3}
b_{u,i} - b_{u',j}
  + (1-\alpha_{u,i,u',j})Y_{\max} + \beta_{u,i,u',j}Y_{\max}
&\geq d^{y,\mathrm{min}}_{u,i,u',j}\\
\label{eq:c1_4}
b_{u',j} - b_{u,i}
  + (1-\alpha_{u,i,u',j})Y_{\max} + (1-\beta_{u,i,u',j})Y_{\max}
&\geq d^{y,\mathrm{min}}_{u,i,u',j}
\end{align}
with
\begin{align}
\label{eq:cqm_dxmin}
d^{x,\mathrm{min}}_{u,i,u',j}
&=
\dfrac{l_{u,i}(1-c_{u,i})+w_{u,i}c_{u,i}}{2}
+
\dfrac{l_{u',j}(1-c_{u',j})+w_{u',j}c_{u',j}}{2}
+ Dx_{u,u'}\\
\label{eq:cqm_dymin}
d^{y,\mathrm{min}}_{u,i,u',j}
&=
\dfrac{l_{u,i}c_{u,i}+w_{u,i}(1-c_{u,i})}{2}
+
\dfrac{l_{u',j}c_{u',j}+w_{u',j}(1-c_{u',j})}{2}
+ Dy_{u,u'}
\end{align}
Here, $\alpha_{u,i,u',j}$ and $\beta_{u,i,u',j}$ determine the
relative position between two rectangles. The four combinations
$(0,0)$, $(0,1)$, $(1,0)$, and $(1,1)$ correspond to the cases in
which rectangle $i$ is placed to the right of, to the left of, below,
and above rectangle $j$, respectively, while satisfying the required
minimum distance in the corresponding direction.
For the area constraint, each rectangle of each functional unit must be
placed within the feasible layout area, as formulated in
Eqs.~\eqref{eq:cqm_area_x} and \eqref{eq:cqm_area_y}.
\begin{align}
\label{eq:cqm_area_x}
\underline{X}_{u,i}
+
\frac{l_{u,i}(1-c_{u,i})+w_{u,i}c_{u,i}}{2}
&\leq
a_{u,i}
\leq
\overline{X}_{u,i}
-
\frac{l_{u,i}(1-c_{u,i})+w_{u,i}c_{u,i}}{2}\\
\label{eq:cqm_area_y}
\underline{Y}_{u,i}
+
\frac{l_{u,i}c_{u,i}+w_{u,i}(1-c_{u,i})}{2}
&\leq
b_{u,i}
\leq
\overline{Y}_{u,i}
-
\frac{l_{u,i}c_{u,i}+w_{u,i}(1-c_{u,i})}{2}
\end{align}
Here, $\underline{X}_{u,i},\overline{X}_{u,i},\underline{Y}_{u,i}$,
and $\overline{Y}_{u,i}$ denote the lower and upper bounds of the
feasible layout area excluding the infeasible placement area and the
transport route area. For a composite functional unit $u$ with two
rectangles $i,j\in R_u$, the combination constraint ensures the
relative arrangement between the two rectangles. First, the two
rectangles share the same rotation state (Eq.~\eqref{eq:cqm_rot}).
\begin{equation}
\label{eq:cqm_rot}
c_{u,i}-c_{u,j}=0
\end{equation}
For the non-mirrored case $\mu_{u}=0$, one binary auxiliary variable
$\gamma_{u,ij}$ is used to determine the feasible relative arrangement
(Eqs.~\eqref{eq:cqm_cnon1} and \eqref{eq:cqm_cnon2}).
\begin{align}
\label{eq:cqm_cnon1}
a_{u,i}-a_{u,j}
&=
(1-c_{u,i})(1-2\gamma_{u,ij})\Delta x_{u,ij}
+
c_{u,i}(2\gamma_{u,ij}-1)\Delta y_{u,ij}\\
\label{eq:cqm_cnon2}
b_{u,i}-b_{u,j}
&=
(1-c_{u,i})(1-2\gamma_{u,ij})\Delta y_{u,ij}
+
c_{u,i}(1-2\gamma_{u,ij})\Delta x_{u,ij}
\end{align}
Here, the binary combinations $(c_{u,i},\gamma_{u,ij})=(0,0)$,
$(1,0)$, $(0,1)$, and $(1,1)$ describe the four possible rotational
states of the composite unit, namely $0^\circ$, $90^\circ$, $180^\circ$,
and $270^\circ$, respectively. For the mirrored cases $\mu_{u}=1$,
two binary auxiliary variables $\gamma^x_{u,ij}$ and
$\gamma^y_{u,ij}$ are used to determine the feasible relative
arrangement considering both rotation and mirroring
(Eqs.~\eqref{eq:cqm_cmir1} and \eqref{eq:cqm_cmir2}).
\begin{align}
\label{eq:cqm_cmir1}
a_{u,i}-a_{u,j}
&=
(2\gamma^x_{u,ij}-1)
\left[
(1-c_{u,i})\Delta x_{u,ij}
+
c_{u,i}\Delta y_{u,ij}
\right]\\
\label{eq:cqm_cmir2}
b_{u,i}-b_{u,j}
&=
(2\gamma^y_{u,ij}-1)
\left[
(1-c_{u,i})\Delta y_{u,ij}
+
c_{u,i}\Delta x_{u,ij}
\right]
\end{align}
For the transport objective, instead of using real-valued distance variables as in prior work \cite{wu2025factory}, auxiliary variables are introduced to represent the Manhattan distances between functional units, and between functional units and the layout input/output.
The corresponding horizontal and vertical distance terms are given in Eqs.~\eqref{eq:cqm_dis_mm}--\eqref{eq:cqm_dis_out}, with the associated constraints defined in Eqs.~\eqref{eq:cqm_dis_mm_con}--\eqref{eq:cqm_dis_out_con}.
{\setlength{\abovedisplayskip}{3pt}
\setlength{\belowdisplayskip}{3pt}
\setlength{\abovedisplayshortskip}{3pt}
\setlength{\belowdisplayshortskip}{3pt}
\begin{equation}
\label{eq:cqm_dis_mm}
\begin{aligned}
d^{x}_{u,i,u',j} &= (2\lambda^{x}_{u,i,u',j}-1)(a_{u,i}-a_{u',j}) \\
d^{y}_{u,i,u',j} &= (2\lambda^{y}_{u,i,u',j}-1)(b_{u,i}-b_{u',j})
\end{aligned}
\end{equation}
\begin{equation}
\label{eq:cqm_dis_in}
\begin{aligned}
d^{x}_{I,u,i} &= (2\lambda^{x}_{I,u,i}-1)(a_{u,i}-x_I) \\
d^{y}_{I,u,i} &= (2\lambda^{y}_{I,u,i}-1)(b_{u,i}-y_I)
\end{aligned}
\end{equation}
\begin{equation}
\label{eq:cqm_dis_out}
\begin{aligned}
d^{x}_{u,i,O} &= (2\lambda^{x}_{u,i,O}-1)(a_{u,i}-x_O) \\
d^{y}_{u,i,O} &= (2\lambda^{y}_{u,i,O}-1)(b_{u,i}-y_O)
\end{aligned}
\end{equation}
with 
\begin{equation}
\label{eq:cqm_dis_mm_con}
\begin{aligned}
(2\lambda^{x}_{u,i,u',j}-1)(a_{u,i}-a_{u',j}) &\geq 0 \\
(2\lambda^{y}_{u,i,u',j}-1)(b_{u,i}-b_{u',j}) &\geq 0
\end{aligned}
\end{equation}
\begin{equation}
\label{eq:cqm_dis_in_con}
\begin{aligned}
(2\lambda^{x}_{I,u,i}-1)(a_{u,i}-x_I) &\geq 0 \\
(2\lambda^{y}_{I,u,i}-1)(b_{u,i}-y_I) &\geq 0
\end{aligned}
\end{equation}
\begin{equation}
\label{eq:cqm_dis_out_con}
\begin{aligned}
(2\lambda^{x}_{u,i,O}-1)(a_{u,i}-x_O) &\geq 0 \\
(2\lambda^{y}_{u,i,O}-1)(b_{u,i}-y_O) &\geq 0
\end{aligned}
\end{equation}
}
Accordingly, the transport objective is formulated in Eq.~\eqref{eq:cqm_obj}.
\begin{equation}
\label{eq:cqm_obj}
\begin{aligned}
H_{\mathrm{CQM}}=\min \Bigl({}&
\sum_{\substack{u,u'\in U\\u<u'}}
\sum_{i\in R_u}\sum_{i'\in R_{u'}}
\left(M_{u,u'}+M_{u',u}\right)
\left(d^{x}_{u,i,u',i'}+d^{y}_{u,i,u',i'}\right)\\
&+\sum_{u\in U}\sum_{i\in R_u}
M_{I,u}\left(d^{x}_{I,u,i}+d^{y}_{I,u,i}\right)
+\sum_{u\in U}\sum_{i\in R_u}
M_{u,O}\left(d^{x}_{u,i,O}+d^{y}_{u,i,O}\right)\Bigr)
\end{aligned}
\end{equation}
\subsubsection{NLM formulation}
Analogous to the CQM formulation, the NLM formulation uses integer coordinate variables and binary orientation variables to define the placement of functional units. Since NLM supports nonlinear modeling and logical operations, the placement of a composite functional unit is determined directly from an anchor rectangle instead of by two rectangles with an additional combination constraint. The decision variables are defined in Eq.~\eqref{eq:nlm_var}.
\begin{equation}
\label{eq:nlm_var}
\begin{aligned}
\hat{a}_{u,i}&\in\{0,1,\ldots, X_{\max}\}\\
\hat{b}_{u,i}&\in\{0,1,\ldots,Y_{\max}\}\\
\hat{c}_{u,i}&\in\{0,1\}
\end{aligned}
\end{equation}
For a composite functional unit $u$ with $R_u=\{i,j\}$ and $i<j$, rectangle $i$ is used as the anchor rectangle. Its center coordinates and orientation are described by $\hat{a}_{u,i}$, $\hat{b}_{u,i}$, and $\hat{c}_{u,i}$, whereas the center coordinates and orientation of rectangle $j$ are derived from the predefined relative arrangement. For the non-mirrored case, the functional unit has four orientations. Eq.~\eqref{eq:nlm_nm_coord} determines the corresponding coordinate relations with an auxiliary binary variable $\hat{\gamma}_{u,ij}$.
\begin{equation}
\label{eq:nlm_nm_coord}
\begin{aligned}
\hat{a}_{u,j}
&=
\hat{a}_{u,i}
-
\left[
(1-\hat{c}_{u,i})(1-2\hat{\gamma}_{u,ij})\Delta x_{u,ij}
+
\hat{c}_{u,i}(2\hat{\gamma}_{u,ij}-1)\Delta y_{u,ij}
\right]\\
\hat{b}_{u,j}
&=
\hat{b}_{u,i}
-
\left[
(1-\hat{c}_{u,i})(1-2\hat{\gamma}_{u,ij})\Delta y_{u,ij}
+
\hat{c}_{u,i}(1-2\hat{\gamma}_{u,ij})\Delta x_{u,ij}
\right]\\
\hat{c}_{u,j}
&=
\hat{c}_{u,i}
\end{aligned}
\end{equation}
Here, the combinations of $\hat{c}_{u,i}$ and $\hat{\gamma}_{u,ij}$ correspond to the four rotations of the composite unit. For the mirrored case, the composite functional unit additionally allows mirrored relative arrangements. The corresponding coordinate relations are defined in Eq.~\eqref{eq:nlm_m_coord} with two auxiliary binary variables $\hat{s}^{x}_{u,ij}$ and $\hat{s}^{y}_{u,ij}$.
\begin{equation}
\label{eq:nlm_m_coord}
\begin{aligned}
\hat{a}_{u,j}
&=
\hat{a}_{u,i}
-
(2\hat{s}^{x}_{u,ij}-1)
\left[
(1-\hat{c}_{u,i})|\Delta x_{u,ij}|
+
\hat{c}_{u,i}|\Delta y_{u,ij}|
\right]\\
\hat{b}_{u,j}
&=
\hat{b}_{u,i}
-
(2\hat{s}^{y}_{u,ij}-1)
\left[
(1-\hat{c}_{u,i})|\Delta y_{u,ij}|
+
\hat{c}_{u,i}|\Delta x_{u,ij}|
\right]\\
\hat{c}_{u,j}
&=
\hat{c}_{u,i}
\end{aligned}
\end{equation}
As in Eq.~\eqref{eq:nlm_nm_coord}, rectangle $i$ denotes the anchor rectangle and rectangle $j$ denotes the derived rectangle. The variable $\hat{c}_{u,i}$ determines the orientation of the composite unit, while $\hat{s}^{x}_{u,ij}$ and $\hat{s}^{y}_{u,ij}$ determine the mirrored relative placement of rectangle $j$.
For the non-overlapping and minimum-distance constraint, a single disjunctive constraint is formulated with absolute value and logical-or operations, as given in Eq.~\eqref{eq:nlm_non_overlap}.
\begin{equation}
\label{eq:nlm_non_overlap}
\begin{aligned}
&|\hat{a}_{u,i}-\hat{a}_{u',i'}|
\geq
\dfrac{l_{u,i}(1-\hat{c}_{u,i})+w_{u,i}\hat{c}_{u,i}}{2}
+
\dfrac{l_{u',i'}(1-\hat{c}_{u',i'})+w_{u',i'}\hat{c}_{u',i'}}{2}
+
Dx_{u,u'}\\
&\vee\;
|\hat{b}_{u,i}-\hat{b}_{u',i'}|
\geq
\dfrac{l_{u,i}\hat{c}_{u,i}+w_{u,i}(1-\hat{c}_{u,i})}{2}
+
\dfrac{l_{u',i'}\hat{c}_{u',i'}+w_{u',i'}(1-\hat{c}_{u',i'})}{2}
+
Dy_{u,u'}
\end{aligned}
\end{equation}
The single disjunctive constraint reduces the need for additional auxiliary variables for direction selection and distance representation. For the area constraint, each rectangle must remain within the admissible layout area. Analogous to the CQM formulation, this is expressed by Eqs.~\eqref{eq:nlm_area_x} and \eqref{eq:nlm_area_y}.
\begin{align}
\label{eq:nlm_area_x}
\underline{X}_{u,i}
+
\frac{l_{u,i}(1-\hat{c}_{u,i})+w_{u,i}\hat{c}_{u,i}}{2}
&\leq
\hat{a}_{u,i}
\leq
\overline{X}_{u,i}
-
\frac{l_{u,i}(1-\hat{c}_{u,i})+w_{u,i}\hat{c}_{u,i}}{2}\\
\label{eq:nlm_area_y}
\underline{Y}_{u,i}
+
\frac{l_{u,i}\hat{c}_{u,i}+w_{u,i}(1-\hat{c}_{u,i})}{2}
&\leq
\hat{b}_{u,i}
\leq
\overline{Y}_{u,i}
-
\frac{l_{u,i}\hat{c}_{u,i}+w_{u,i}(1-\hat{c}_{u,i})}{2}
\end{align}
Here, $\underline{X}_{u,i}$, $\overline{X}_{u,i}$, $\underline{Y}_{u,i}$, and $\overline{Y}_{u,i}$ denote the lower and upper bounds of the feasible layout area. In addition, rectangles must not overlap with the transport route area $TP$ or the infeasible placement area $IM$, which can be represented as a set of forbidden rectangles. For a forbidden rectangular area $q\in TP\cup IM$ with center coordinates $(x_q,y_q)$, and dimensions $l_q$ and $w_q$, a single disjunctive constraint is formulated in Eq.~\eqref{eq:nlm_area_forbid}.
\begin{equation}
\label{eq:nlm_area_forbid}
|\hat{a}_{u,i}-x_q|
\geq
\frac{l_{u,i}(1-\hat{c}_{u,i})+w_{u,i}\hat{c}_{u,i}}{2}
+
\frac{l_q}{2}
\;\;\vee\;\;
|\hat{b}_{u,i}-y_q|
\geq
\frac{l_{u,i}\hat{c}_{u,i}+w_{u,i}(1-\hat{c}_{u,i})}{2}
+
\frac{w_q}{2}
\end{equation}

The objective of the NLM formulation is to minimize transport intensity. Let $\bar{i}_u\in R_u$ denote the rectangle whose center represents the transport center of functional unit $u$. The transport objective is then formulated in Eq.~\eqref{eq:nlm_obj}.
\begin{equation}
\label{eq:nlm_obj}
\begin{aligned}
H_{\mathrm{NLM}}=\min \Bigg(
&\sum_{\substack{u,u'\in U\\u<u'}}
\left(M_{u,u'}+M_{u',u}\right)
\left(
|\hat{a}_{u,\bar{i}_u}-\hat{a}_{u',\bar{i}_{u'}}|
+
|\hat{b}_{u,\bar{i}_u}-\hat{b}_{u',\bar{i}_{u'}}|
\right)\\
&+\sum_{u\in U}
M_{I,u}
\left(
|\hat{a}_{u,\bar{i}_u}-x_I|
+
|\hat{b}_{u,\bar{i}_u}-y_I|
\right)
+\sum_{u\in U}
M_{u,O}
\left(
|\hat{a}_{u,\bar{i}_u}-x_O|
+
|\hat{b}_{u,\bar{i}_u}-y_O|
\right)
\Bigg)
\end{aligned}
\end{equation}
Accordingly, the transport intensity is represented directly by the Manhattan distances between the corresponding transport centers.

\subsection{Iterative approach}
\label{subsec:iterative}

Although the D-Wave Hybrid Solver applied in IBQM already has a decomposition mechanism, an additional iterative approach is developed to decompose the problem into subproblems according to placement priority. This further reduces the number of variables in each optimization step and increases the scalability of IBQM for large layout instances. The complete procedure is presented in Algorithm~\ref{alg:iterative}.

\begin{algorithm}[t]
\caption{Iterative BQM-based factory layout planning}
\label{alg:iterative}
\begin{algorithmic}[1]
\STATE \textbf{Input:} Layout area, functional units $U$, material flow, distance requirements, the number of functional units per iteration $n_{\text{sub}}$, initial area ratios $\rho_x^{(0)},\rho_y^{(0)}$, area expansion speeds $\eta_{x},\eta_{y}$
\STATE \textbf{Output:} Optimized layout solution
\STATE \textbf{Initialization:} $\text{loop} \gets 0$, \; $U_{\mathrm{rem}} \gets U$, \; $U_{\mathrm{placed}} \gets \emptyset$
\WHILE{$U_{\mathrm{rem}}\neq\emptyset$}
    \STATE Determine the factor of transport flow $\mathcal{F}_{ts,u}^{(\mathrm{loop})}$ for each $u\in U_{\mathrm{rem}}$
    \STATE Select seed $u^* \gets \arg\max_{u \in U_{\mathrm{rem}}}\mathcal{F}_{ts,u}^{(\mathrm{loop})}$
    \STATE $U_{\mathrm{sub}}^{(\mathrm{loop})} \gets \{u^*\}$
    \WHILE{$|U_{\mathrm{sub}}^{(\mathrm{loop})}| < n_{\text{sub}}$ \textbf{and} $U_{\mathrm{rem}} \setminus U_{\mathrm{sub}}^{(\mathrm{loop})} \neq \emptyset$}
        \FOR{each $u \in U_{\mathrm{rem}} \setminus U^{(\mathrm{loop})}_{\mathrm{sub}}$}
            \STATE Identify the factor of connectivity $\mathcal{F}_{co,u}^{(\mathrm{loop})}$
        \ENDFOR
        \STATE $B_{u}^{(\mathrm{loop})}\gets 0.7\,\mathcal{F}_{ts,u}^{(\mathrm{loop})} + 0.3\,\mathcal{F}_{co,u}^{(\mathrm{loop})}$
        \STATE $U^{(\mathrm{loop})}_{\mathrm{sub}}\gets U^{(\mathrm{loop})}_{\mathrm{sub}}\cup\{\arg\max_{u \in U_{\mathrm{rem}} \setminus U_{\mathrm{sub}}^{(\mathrm{loop})}}B_u^{(\mathrm{loop})}\}$
    \ENDWHILE
    \STATE Determine the search-area reference point $(x_{\mathrm{ref}}^{(\mathrm{loop})},y_{\mathrm{ref}}^{(\mathrm{loop})})$
    \STATE Identify the search layout area $A^{(\mathrm{loop})}$ from the area ratios and the reference point
    \STATE Generate candidate positions $P^{(\mathrm{loop})}_{u,i}$ for each rectangle $i \in R_u$, $u \in U_{\mathrm{sub}}^{(\mathrm{loop})}$ within $A^{(\mathrm{loop})}$
    \STATE Build BQM
    \STATE Solve BQM
    \IF{any constraint violated}
        \STATE Increase the corresponding weight and reattempt
    \ENDIF
    \STATE Integrate the subproblem solution into the overall solution
    \STATE Update $x_{\mathrm{ref}}^{(\mathrm{loop}+1)}$, $y_{\mathrm{ref}}^{(\mathrm{loop}+1)}$
    \STATE $\rho_x^{(\mathrm{loop}+1)}\gets \min\{1,\rho_x^{(\mathrm{loop})}+\eta_x(1-\rho_x^{(\mathrm{loop})})\}$
    \STATE $\rho_y^{(\mathrm{loop}+1)}\gets \min\{1,\rho_y^{(\mathrm{loop})}+\eta_y(1-\rho_y^{(\mathrm{loop})})\}$
    \STATE $U_{\mathrm{rem}}\gets U_{\mathrm{rem}}\setminus U_{\mathrm{sub}}^{(\mathrm{loop})}$
    \STATE $\mathrm{loop}\gets \mathrm{loop}+1$
\ENDWHILE
\RETURN Overall solution
\end{algorithmic}
\end{algorithm}
The iterative approach focuses on the definition of the subproblem through the determination of the limited layout area and the subset of functional units. The optimized layout solution is gradually created by solving the BQM of each subproblem. The process begins with the definition of the subset of the functional units. According to the considered objective of transport intensity, the selection is at first related to the normalized transport flow factor $\mathcal{F}_{ts,u}^{(\mathrm{loop})}$, which is given by Eq.~\eqref{eq:factor_ts}. The material flow considers the flow between the functional unit and other remaining functional units, the flow between the functional unit and the layout input and output, and the flow between the functional unit and already-placed units. Since the placement of a functional unit with strong material flow connections to placed units has a direct influence on the transport intensity of the partial solution, the flow to placed units is weighted with a factor of 2.
\begin{equation}
\label{eq:factor_ts}
\mathcal{F}^{(\mathrm{loop})}_{ts,u}= \frac{F^{(\mathrm{loop})}_{ts,u} - \min_{u'\in U_{\mathrm{rem}}}F^{(\mathrm{loop})}_{ts,u'}}{\max_{u'\in U_{\mathrm{rem}}}F^{(\mathrm{loop})}_{ts,u'} -\min_{u'\in U_{\mathrm{rem}}}F^{(\mathrm{loop})}_{ts,u'}}
\end{equation}
with
\begin{equation}
\label{eq:Fts}
F_{ts,u}^{(\mathrm{loop})}=\sum_{\substack{u'\in U_{\mathrm{rem}}\\u'\neq u}}M_{u,u'}+M_{I,u}+M_{u,O}+2\sum_{u'\in U\setminus U_{\mathrm{rem}}} M_{u,u'}
\end{equation}
The functional unit with the largest transport flow factor is selected as the seed in the subset. Based on this seed, subsequent units are selected iteratively by a weighted combination of the transport flow factor and the connectivity factor. The connectivity factor quantifies the transport interaction between the candidate unit and the already-selected units in the subset and is given by Eq.~\eqref{eq:factor_co}.
\begin{equation}
\label{eq:factor_co}
\mathcal{F}_{co,u}^{(\mathrm{loop})} = \frac{F^{(\mathrm{loop})}_{co,u} - \min_{u'\in U_{\mathrm{rem}}}F^{(\mathrm{loop})}_{co,u'}}{\max_{u'\in U_{\mathrm{rem}}}F^{(\mathrm{loop})}_{co,u'} -\min_{u'\in U_{\mathrm{rem}}}F^{(\mathrm{loop})}_{co,u'}}
\end{equation}
with
\begin{equation}
\label{eq:Fco}
F^{(\mathrm{loop})}_{co,u}=\sum_{u' \in U_{\mathrm{sub}}^{(\mathrm{loop})}} (M_{u,u'} + M_{u',u})
\end{equation}

After the definition of the subset, the layout area is limited. Its size is controlled by the area ratios $\rho_x^{(\mathrm{loop})}$ and $\rho_y^{(\mathrm{loop})}$, which describe the fraction of the full layout length and width considered in the current iteration. Starting from the initial ratios $\rho_x^{(0)}$ and $\rho_y^{(0)}$, the ratios are updated by the area expansion speeds $\eta_{x},\eta_{y}$ in each iteration.
The search area is centered around the reference point $(x_{\mathrm{ref}}^{(\mathrm{loop})},y_{\mathrm{ref}}^{(\mathrm{loop})})$. In the first iteration, it is set to the centroid of the transport corridor. In subsequent iterations, it is updated to the centroid of the placed functional units, as given by Eq.~\eqref{eq:ref_point}.
\begin{equation}
\label{eq:ref_point}
x_{\mathrm{ref}}^{(\mathrm{loop})} = \frac{1}{|U_{\mathrm{placed}}|}\sum_{u \in U_{\mathrm{placed}}} x_u, \qquad
y_{\mathrm{ref}}^{(\mathrm{loop})} = \frac{1}{|U_{\mathrm{placed}}|}\sum_{u \in U_{\mathrm{placed}}} y_u
\end{equation}
With the spans $s_x^{(\mathrm{loop})}=\rho_x^{(\mathrm{loop})}X_{\max}$ and $s_y^{(\mathrm{loop})}=\rho_y^{(\mathrm{loop})}Y_{\max}$, the search area $A^{(\mathrm{loop})}$ is defined by Eq.~\eqref{eq:search_area}, where the lower corner is clamped to keep $A^{(\mathrm{loop})}$ within the layout domain.
\begin{equation}
\label{eq:search_area}
A^{(\mathrm{loop})} = \bigl[x_{\min}^{(\mathrm{loop})},\; x_{\min}^{(\mathrm{loop})} + s_x^{(\mathrm{loop})}\bigr] \times \bigl[y_{\min}^{(\mathrm{loop})},\; y_{\min}^{(\mathrm{loop})} + s_y^{(\mathrm{loop})}\bigr]
\end{equation}
with
\begin{equation}
\label{eq:search_clamp}
\begin{aligned}
x_{\min}^{(\mathrm{loop})} &= \min\!\left\{\max\!\left\{x_{\mathrm{ref}}^{(\mathrm{loop})}-\tfrac{s_x^{(\mathrm{loop})}}{2},\; 0\right\},\; X_{\max}-s_x^{(\mathrm{loop})}\right\} \\
y_{\min}^{(\mathrm{loop})} &= \min\!\left\{\max\!\left\{y_{\mathrm{ref}}^{(\mathrm{loop})}-\tfrac{s_y^{(\mathrm{loop})}}{2},\; 0\right\},\; Y_{\max}-s_y^{(\mathrm{loop})}\right\}
\end{aligned}
\end{equation}

Within $A^{(\mathrm{loop})}$, the candidate positions for each rectangle of the selected functional units are generated on the discretized coordinate grid. Candidates overlapping with the transport route area, the infeasible placement area, or the occupied area of an already-placed functional unit are removed. The BQM is then formulated for the selected subset with the placed units treated as fixed obstacles and solved. If a constraint is not fulfilled, the computation will be restarted with the increasing corresponding constraint weight. After solving the current subproblem, the accepted placements are integrated into the overall layout solution and registered as fixed placements for the subsequent iteration. The remaining set $U_{rem}$ is updated accordingly, and both the reference point and the area ratios are renewed for the next iteration. The same procedure is repeated until all functional units have been placed.
\subsection{Variable pruning}
\label{subsec:variable_pruning} 
To improve the scalability of QA-based approaches, variable pruning is used to reduce unnecessary decision variables and restrict the search space. For IBQM, binary variables corresponding to infeasible candidate positions are eliminated. These infeasible candidates include positions that overlap with the transport route area or infeasible placement area, as well as positions that violate the minimum-distance requirements with already placed functional units from previous iterations. Therefore, the candidate positions of rectangle $i\in R_u$ in iteration $\mathrm{loop}$ are given by Eq.~\eqref{eq:candi_ps_iter_pruned}.
\begin{equation}
\label{eq:candi_ps_iter_pruned}
\begin{aligned}
P_{u,i}^{(\mathrm{loop})}
=
\bigg\{
p=(x,y,k)\;\bigg|\;&
 k\in\{0,1\},\;
(x,y)\in A^{(\mathrm{loop})},\;p\cap TP=\varnothing,\;
p\cap IM=\varnothing,\;\forall u'\in \mathcal{U}_{\mathrm{placed}}^{(\mathrm{loop})},j\in R_{u'}\\
&\quad
\left(
|x-x^{*}_{u',j}|
-\frac{l_{u,i}^{k}}{2}
-\frac{l_{u',j}^{k^{*}_{u',j}}}{2}
\geq Dx_{u,u'}
\right) \vee
\left(
|y-y^{*}_{u',j}|
-\frac{w_{u,i}^{k}}{2}
-\frac{w_{u',j}^{k^{*}_{u',j}}}{2}
\geq Dy_{u,u'}
\right)
\bigg\}
\end{aligned}
\end{equation}
In HDQM, scalability is affected by both the number of discrete variables and the number of cases assigned to each variable. The variable pruning for HDQM removes infeasible placement candidates. In addition, for large problem instances, a flow-distance-based candidate selection is used to limit the number of cases assigned to each discrete variable. The selection is based on a case selection score related to the transport route area, the input/output material flow, and the material-flow relations with other functional units. For each functional unit $u$, the selected case set is given by Eq.~\eqref{eq:dqm_case_selection}.
\begin{equation}
\label{eq:dqm_case_selection}
S_u^{\mathrm{sel}}
=
\operatorname*{arg\,min}_{\substack{\mathcal{S}\subseteq S_u\\|\mathcal{S}|=\min\{|S_u|,N_s\}}}
\sum_{s\in \mathcal{S}}
\Psi_{u,s}
\end{equation}
where $N_s$ denotes the maximum number of cases assigned to one discrete variable, and $\Psi_{u,s}$ denotes the criterion value of case $s\in S_u$. The selection criterion considers the distance to the transport route area, the input/output material flow, and the material-flow relations with other functional units. The criterion value is given by Eq.~\eqref{eq:dqm_case_score}.
\begin{equation}
\label{eq:dqm_case_score}
\Psi_{u,s}
=
\delta^{TP}_{u,s}
+
\delta^{IO}_{u,s}
+
\delta^{UU}_{u,s}
\end{equation}
with
\begin{equation}
\label{eq:dqm_case_score_terms}
\begin{aligned}
\delta^{TP}_{u,s}
&=
\frac{1}{|R_u|}
\sum_{i\in R_u}
d^{TP}_{u,i,s}\\[3pt]
\delta^{IO}_{u,s}
&=
\frac{
M_{I,u}d^{I}_{u,s}
+
M_{u,O}d^{O}_{u,s}
}{
M_{I,u}+M_{u,O}
}\\[3pt]
\delta^{UU}_{u,s}
&=
\frac{
\sum_{\substack{u'\in U\\u'\neq u}}
(M_{u,u'}+M_{u',u})
d^{UU}_{u,u',s}
}{
\sum_{\substack{u'\in U\\u'\neq u}}
(M_{u,u'}+M_{u',u})
}
\end{aligned}
\end{equation}
where $d^{TP}_{u,i,s}$, $d^{I}_{u,s}$, $d^{O}_{u,s}$, and $d^{UU}_{u,u',s}$ are given in Eqs.~\eqref{eq:app_dqm_tp_distance}--\eqref{eq:app_dqm_uu_distance}.
For HCQM and HNLM, the search space is limited by defining the domains of the coordinate variables only over feasible placement ranges.
\section{Numerical case study}
\label{sec:results}
\subsection{Experimental setup and parameter setting}
\label{subsec:instances}
The experimental setup is designed to evaluate the applicability and scalability of different QA-based formulations under layout conditions with increasing structural complexity. The test instances are derived from the industrial use case considered in previous work \cite{wu2025factory} and are extended into five experimental settings that vary in problem scale, variability in unit dimensions, geometric composition, spatial restrictions, and the configuration of transport routes.
The five experimental settings are summarized in Table~\ref{tab:benchmark_instances}. In Setting A, all functional units are single rectangular units with identical dimensions within each instance. From A1 to A4, the number of functional units $n$ increases from 10 to 75, while the layout size and unit dimensions are scaled accordingly. This setting is designed to analyze scale effects under homogeneous unit geometry.  In Setting B, all functional units are represented by single rectangles with randomly sampled dimensions. The instances use square layout areas and increase in scale from B1 to B4. The variability in unit length and width leads to less regular spatial arrangements and increases the diversity of distance relations between functional units.

\begin{figure}[pos=t]
\centering
\includegraphics[width=.9\linewidth]{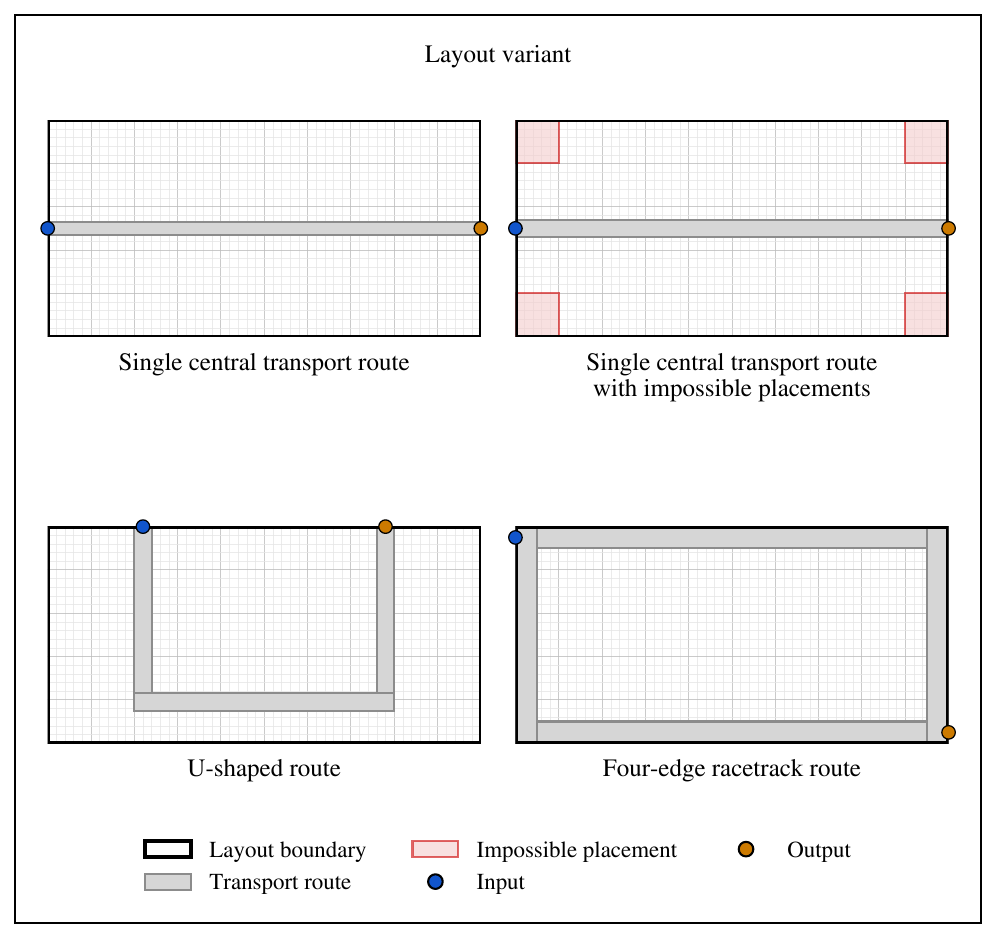}
\caption{Layout variants with different transport routes.}
\label{fig:layout_variants}
\end{figure}

Setting C increases the geometric complexity of the layout problem by combining rectangular and composite functional units. The inclusion of composite units expands the orientation decision space and adds combination constraints. Setting D represents brownfield layout conditions by considering infeasible placement areas that represent existing equipment or infrastructure. In Settings A to D, the transport structure is kept as a central straight route, while Setting E considers practical transport route configurations. In particular, E1 uses a U shaped transport route, whereas E2 uses a racetrack transport route with a cross aisle. The four layout variants considered in the experimental settings are illustrated in Fig.~\ref{fig:layout_variants}.
For all instances, material flow values and minimum distance requirements are specified at the instance level, and the detailed instance data are provided in the Supplementary Material.
\begin{table}[width=\linewidth,cols=6,pos=t]
\caption{Benchmark scenarios used in the experimental evaluation.}
\label{tab:benchmark_instances}
\small
\renewcommand{\arraystretch}{1.25}
\setlength{\tabcolsep}{5pt}
\begin{tabular*}{\tblwidth}{@{}LLLLLL@{}}
\toprule
Experiment setting & Instances & \makecell[l]{Unit\\geometry} & $n$ & \makecell[l]{Layout\\size} & \makecell[l]{Layout\\characteristics} \\
\midrule
A & A1--A4 & \makecell[l]{Identical\\rectangles} & 10--75 & $10{\times}10$--$100{\times}150$ & Central straight route \\
B & B1--B4 & \makecell[l]{Random\\rectangles} & 10--75 & $25{\times}25$--$150{\times}150$ & Central straight route \\
C & C1--C2 & \makecell[l]{Rectangles and\\composite units} & 25--50 & $50{\times}100$--$150{\times}200$ & Central straight route \\
D & D1--D2 & \makecell[l]{Rectangles and\\composite units} & 20--50 & $50{\times}100$--$150{\times}200$ & \makecell[l]{Central straight route with\\corner obstacles} \\
E & E1--E2 & \makecell[l]{Rectangles and\\composite units} & 20--50 & $50{\times}100$--$150{\times}200$ & \makecell[l]{U-shaped route; racetrack\\route with cross-aisle} \\
\bottomrule
\end{tabular*}
\end{table}
Each instance is solved using four QA-based approaches, namely IBQM, HDQM, HCQM, and HNLM, and three classical metaheuristic approaches, namely GA, PSO, and TBSA, where TBSA refers to a hybrid TB and SA method. Since the D-Wave hybrid BQM solver used in IBQM combines TB, SA, and QA, TBSA is implemented with the same BQM input and the same iterative approach as IBQM. The comparison between TBSA and IBQM is used to further evaluate the effect of QA.
The parameter settings are specified according to the size and complexity of the corresponding instances. Therefore, for GA and PSO, a full-factorial parameter study is conducted to identify suitable parameter settings for each instance.
For HDQM, HCQM, and HNLM, the internal search procedures of the Leap hybrid solvers are managed by D-Wave and are not directly configurable. 
Therefore, the solver time limit is used to control the computation time of the corresponding solver. 
The solver time limits for HDQM, HCQM, and HNLM and the parameter settings for GA and PSO are summarized in Table~\ref{tab:parameters_direct}.

Due to the decomposition structure of IBQM and TBSA, the subproblems solved in each loop have a comparable size. 
Therefore, fixed parameter settings are used for all instances. 
In each iteration, the SA phase is executed with 25 reads and 10,000 sweeps per read, while the TB phase is executed with 25 reads. For IBQM, the QA phase is additionally configured with 300 QPU reads.
All computations are performed on an Intel Core(TM) i7-1165G7 processor with 16 GB RAM. The quantum resource for IBQM is the D-Wave Advantage2 System 1 with Zephyr topology. HDQM, HCQM, and HNLM are implemented using the corresponding Leap hybrid solvers from D-Wave.
\begin{table}[width=\linewidth,cols=13,pos=t]
\caption{Parameter settings for the direct solution approaches.}
\label{tab:parameters_direct}
\tiny
\renewcommand{\arraystretch}{1.05}
\setlength{\tabcolsep}{1pt}
\begin{tabular*}{\tblwidth}{@{}LLLLLLLLLLLLL@{}}
\toprule
\multirow{2}{*}{Instance}
& \multicolumn{3}{c}{\makecell[c]{Solver time\\limit (s)}}
& \multicolumn{4}{c}{GA}
& \multicolumn{5}{c}{PSO} \\
\cmidrule(lr){2-4}
\cmidrule(lr){5-8}
\cmidrule(lr){9-13}
& HDQM & HCQM & HNLM
& \makecell[l]{Population\\size}
& \makecell[l]{Maximum\\generations}
& \makecell[l]{Crossover\\rate}
& \makecell[l]{Mutation\\rate}
& \makecell[l]{Swarm\\size}
& \makecell[l]{Maximum\\iterations}
& \makecell[l]{Inertia\\weight}
& \makecell[l]{Cognitive\\coefficient}
& \makecell[l]{Social\\coefficient} \\
\midrule
A1 & 10  & 15  & 10  & 80  & 120 & 0.80 & 0.18 & 60  & 160 & 0.75 & 1.40 & 1.70 \\
A2 & 20  & 15  & 20  & 60  & 240 & 0.85 & 0.12 & 70  & 180 & 0.75 & 1.60 & 1.30 \\
A3 & 30  & 70  & 30  & 140 & 300 & 0.75 & 0.03 & 120 & 180 & 0.50 & 1.50 & 1.80 \\
A4 & 60  & 120 & 45  & 180 & 320 & 0.65 & 0.02 & 120 & 280 & 0.40 & 2.20 & 1.90 \\
B1 & 10  & 10  & 10  & 100 & 180 & 0.85 & 0.18 & 50  & 140 & 0.55 & 1.10 & 1.50 \\
B2 & 30  & 10  & 10  & 100 & 200 & 0.75 & 0.11 & 110 & 120 & 0.65 & 1.50 & 1.70 \\
B3 & 30  & 30  & 30  & 160 & 280 & 0.70 & 0.04 & 120 & 220 & 0.60 & 1.90 & 1.50 \\
B4 & 120 & 120 & 30  & 160 & 360 & 0.60 & 0.03 & 120 & 340 & 0.70 & 2.10 & 1.90 \\
C1 & 30  & 30  & 20  & 140  & 300  & 0.85   & 0.08   & 120  & 180  & 0.70   & 1.10   & 1.80   \\
C2 & 120 & 130 & 60  & 160  & 400  & 0.85   & 0.03   & 120  & 320  & 0.65   & 1.00   & 1.50   \\
D1 & 10  & 10  & 10  & 160 & 220 & 0.90 & 0.11 & 110 & 320 & 0.45 & 1.30 & 2.10 \\
D2 & 80  & 120 & 120 & 120 & 220 & 0.85 & 0.07 & 120 & 180 & 0.40 & 2.20 & 1.90 \\
E1 & 10  & 20  & 30  & 80  & 160 & 0.85 & 0.12 & 50  & 140 & 0.70 & 1.50 & 2.30 \\
E2 & 120 & 120 & 120 & 200 & 340 & 0.80 & 0.02 & 120  & 320  & 0.35   & 1.40   & 1.60   \\
\bottomrule
\end{tabular*}
\end{table}

\subsection{Results and comparative analysis}
\label{subsec:rest}
\begin{table}[width=\linewidth,cols=15,pos=t]
\caption{Scale comparison of different QA-based approaches for the evaluated layout instances.}
\label{tab:model_scale_comparison}
\tiny
\renewcommand{\arraystretch}{1.05}
\setlength{\tabcolsep}{1pt}
\begin{tabular*}{\tblwidth}{@{}LLLLLLLLLLLLLLL@{}}
\toprule
\multirow{2}{*}{Instance} &
\multirow{2}{*}{\makecell[c]{Layout\\size}} &
\multirow{2}{*}{$n$} &
\multicolumn{2}{c}{IBQM} &
\multicolumn{3}{c}{HDQM} &
\multicolumn{3}{c}{HCQM} &
\multicolumn{4}{c}{HNLM} \\
\cmidrule(lr){4-5} \cmidrule(lr){6-8} \cmidrule(lr){9-11} \cmidrule(l){12-15}
 & & & Variables & Quadratic & Variables & Quadratic & Cases & Variables & Quadratic & Constraints & Variables & Constraints & \makecell[l]{Decision state\\size} & Nodes \\
\midrule
A1 & $10{\times}10$ & 10 & 1,260 & 298,760 & 10 & 648,000 & 1,200 & 285 & 390 & 435 & 30 & 95 & 240 & 1,811 \\
A2 & $25{\times}50$ & 25 & 12,757 & 8,330,713 & 25 & 4,289,232 & 3,000 & 1,463 & 2,464 & 2,213 & 75 & 425 & 600 & 8,929 \\
A3 & $50{\times}100$ & 50 & 20,000 & 9,186,428 & 50 & 17,556,337 & 6,000 & 5,426 & 9,920 & 8,176 & 150 & 1,475 & 1,200 & 31,783 \\
A4 & $100{\times}150$ & 75 & 30,000 & 11,723,747 & 75 & 109,961,902 & 15,000 & 11,848 & 22,210 & 17,848 & 225 & 3,150 & 1,800 & 68,137 \\
B1 & $25{\times}25$ & 10 & 2,709 & 2,050,938 & 10 & 648,000 & 1,200 & 285 & 390 & 435 & 30 & 95 & 240 & 1,809 \\
B2 & $50{\times}50$ & 25 & 6,527 & 5,383,142 & 25 & 4,288,893 & 3,000 & 1,463 & 2,464 & 2,213 & 75 & 425 & 600 & 6,480 \\
B3 & $100{\times}100$ & 50 & 15,419 & 14,138,032 & 50 & 48,745,250 & 10,000 & 5,426 & 9,920 & 8,176 & 150 & 1,475 & 1,200 & 23,780 \\
B4 & $150{\times}150$ & 75 & 19,592 & 15,943,892 & 75 & 109,935,080 & 15,000 & 11,848 & 22,210 & 17,848 & 225 & 3,150 & 1,800 & 51,536 \\
C1 & $50{\times}100$ & 25 & 10,640 & 6,401,343 & 25 & 4,291,998 & 3,000 & 2,373 & 2,490 & 4,007 & 88 & 880 & 704 & 13,698 \\
C2 & $150{\times}200$ & 50 & 23,649 & 28,291,157 & 50 & 61,484,908 & 17,500 & 9,704 & 9,982 & 16,670 & 181 & 3,614 & 1,448 & 54,757 \\
D1 & $50{\times}100$ & 20 & 10,062 & 5,275,240 & 20 & 2,727,652 & 2,400 & 2,199 & 1,580 & 3,765 & 69 & 658 & 552 & 9,869 \\
D2 & $150{\times}200$ & 50 & 27,013 & 15,861,662 & 50 & 17,562,576 & 6,000 & 9,992 & 9,964 & 16,976 & 172 & 3,182 & 1,376 & 50,174 \\
E1 & $50{\times}100$ & 20 & 8,819 & 5,575,211 & 20 & 2,728,202 & 2,400 & 1,824 & 1,598 & 3,100 & 69 & 600 & 552 & 9,455 \\
E2 & $150{\times}200$ & 50 & 21,070 & 9,938,878 & 50 & 48,753,073 & 10,000 & 10,434 & 9,786 & 17,934 & 175 & 3,425 & 1,400 & 53,892 \\
\bottomrule
\end{tabular*}
\end{table}

\begin{table*}[width=\textwidth,cols=15,pos=t]
\caption{Transport intensity and computation time of all approaches.}
\label{tab:results_ts_time}
\tiny
\renewcommand{\arraystretch}{1.05}
\setlength{\tabcolsep}{1pt}
\begin{tabular*}{\tblwidth}{@{}LLLLLLLLLLLLLLL@{}}
\toprule
\multirow{2}{*}{Instance} &
\multicolumn{2}{c}{GA} &
\multicolumn{2}{c}{PSO} &
\multicolumn{2}{c}{IBQM} &
\multicolumn{2}{c}{TBSA} &
\multicolumn{2}{c}{HCQM} &
\multicolumn{2}{c}{HDQM} &
\multicolumn{2}{c}{HNLM} \\
\cmidrule(lr){2-3} \cmidrule(lr){4-5} \cmidrule(lr){6-7} \cmidrule(lr){8-9} \cmidrule(lr){10-11} \cmidrule(lr){12-13} \cmidrule(lr){14-15}
 & TI & \makecell[l]{Computation\\time in s} & TI & \makecell[l]{Computation\\time in s} & TI & \makecell[l]{Computation\\time in s} & TI & \makecell[l]{Computation\\time in s} & TI & \makecell[l]{Computation\\time in s} & TI & \makecell[l]{Computation\\time in s} & TI & \makecell[l]{Computation\\time in s} \\
\midrule
A1 & 306,378 & 313.47 & 302,434 & 92.63 & \textbf{302,133} & 28.06 & 305,883 & 28.23 & 309,915 & 23.85 & \textbf{302,133} & 11.89 & 314,619 & \textbf{11.51} \\
A2 & 6,516,624 & 36.57 & 7,504,848 & 245.71 & \textbf{6,355,120} & 151.80 & 6,460,112 & 291.00 & 6,783,905 & 28.35 & 6,459,814 & 29.58 & 6,687,573 & \textbf{27.26} \\
A3 & 52,487,073 & 336.51 & 58,748,825 & 2,292.14 & \textbf{48,520,210} & 427.72 & 50,809,491 & 1,082.46 & 54,641,112 & 169.49 & 55,298,106 & 141.67 & 53,552,093 & \textbf{41.99} \\
A4 & 175,315,364 & 469.41 & 192,559,142 & 2,561.45 & 167,620,309 & 799.49 & \textbf{167,386,496} & 1,136.71 & 184,553,613 & 144.33 & 198,130,494 & 156.76 & 178,050,860 & \textbf{79.96} \\
B1 & 661,036 & 14.25 & 686,514 & \textbf{13.98} & 689,674 & 69.42 & 689,848 & 100.31 & 673,682 & 17.01 & 668,732 & 19.26 & \textbf{660,602} & 15.93 \\
B2 & 10,115,269 & 167.61 & 10,423,020 & 1,290.71 & \textbf{9,865,458} & 145.95 & 9,989,994 & 166.06 & 10,498,704 & 20.31 & 10,353,965 & 57.74 & 10,276,773 & \textbf{16.28} \\
B3 & 67,108,408 & 338.08 & 75,204,997 & 5,631.17 & \textbf{64,183,517} & 285.38 & 64,702,480 & 679.66 & 69,599,990 & 138.29 & 67,245,929 & 153.56 & 68,379,618 & \textbf{37.43} \\
B4 & 235,889,267 & 1,235.89 & 254,633,293 & 15,356.71 & \textbf{233,352,067} & 366.47 & 233,907,837 & 643.49 & 256,925,731 & 168.50 & 249,831,529 & 173.32 & 241,955,953 & \textbf{41.97} \\
C1 & 12,502,477 & 340.33 & 13,910,484 & 1,165.09 & \textbf{12,417,458} & 67.34 & 12,905,145 & 147.36 & 13,165,117 & \textbf{44.42} & 12,732,082 & 78.58 & 12,628,944 & 45.97 \\
C2 & \textbf{127,241,985} & 21,866.68 & 141,659,149 & 264,103.37 & 127,977,423 & 450.53 & 127,246,239 & 1,278.00 & 140,931,064 & 166.06 & 140,540,649 & 99.00 & 132,791,292 & \textbf{70.14} \\
D1 & 8,715,853 & 2,115.08 & 9,149,579 & 8,327.40 & \textbf{8,625,067} & 297.00 & 8,762,943 & 379.33 & 9,849,094 & \textbf{22.09} & 9,481,550 & 33.28 & 9,190,171 & 36.37 \\
D2 & 126,494,713 & 102,047.75 & 136,292,412 & 124,596.32 & \textbf{125,861,158} & 559.17 & 127,782,231 & 1,119.96 & 134,565,689 & 170.45 & 142,594,486 & 137.15 & 127,083,248 & \textbf{127.00} \\
E1 & 9,506,254 & 636.79 & 9,607,944 & 2,099.57 & \textbf{9,399,186} & 146.00 & 9,413,438 & 248.52 & 10,646,435 & 38.48 & 9,534,821 & \textbf{34.26} & 9,779,401 & 42.79 \\
E2 & 112,152,275 & 3,190.62 & 124,466,906 & 17,306.22 & \textbf{109,261,013} & 431.00 & 111,193,804 & 966.37 & 116,877,318 & 161.16 & 120,590,744 & 160.86 & 112,964,336 & \textbf{130.43} \\
\bottomrule
\end{tabular*}
\end{table*}

The performance of the implemented approaches is evaluated with respect to scalability, solution quality, and computation time. Table~\ref{tab:model_scale_comparison} reports the solver-specific scale indicators of the QA-based approaches for each instance, which are used to evaluate scalability. For IBQM, the number of variables and quadratic terms corresponds to the accumulated model size over all iterations. Although the decomposition structure enables IBQM to solve all large-scale instances, the accumulated number of quadratic terms still reaches more than $28$ million in C2. This indicates that the scalability of IBQM is limited by the substantial increase in quadratic interactions, especially for layout instances with increasing geometric complexity.

HDQM uses only one discrete variable for each functional unit, so the number of variables remains comparatively small. However, the number of cases and the corresponding quadratic interactions between case pairs dominate scalability. Although the number of cases is reduced by the proposed candidate selection procedure, HDQM exceeds $100$ million quadratic interactions in A4 and B4. Compared with IBQM and HDQM, HCQM and HNLM show stronger scalability. For HCQM, the number of constraints remains below $18{,}000$ even for the large-scale instances, which is well below the corresponding solver limitation. Since HNLM supports nonlinear decision structures and logic operations, the solver-specific scale indicators of HNLM remain comparatively small even when the number of functional units and layout size increases. Accordingly, HCQM and HNLM show greater potential for solving more complex practical layout problems.

The solution quality and computation time of all approaches obtained under the predefined computational settings are summarized in Table~\ref{tab:results_ts_time}. 
Solution quality is evaluated by the transport intensity $TI$ of the feasible layout solution, for which all constraints are satisfied.  Computation time is the total time from the start of the solution procedure to the generation of the final solution. Since the QA-based approaches are implemented through a cloud-based platform, the computation time includes model generation, model submission, solver execution, and result transfer to the local computer. The minimum transport intensity and the shortest computation time for each instance are highlighted in bold.

In terms of solution quality, IBQM achieves the lowest transport intensity in $78.6\%$ of the instances, with a maximum reduction of $17.41\%$ relative to PSO and a corresponding $81.34\%$ reduction in computation time. In the remaining instances, the deviation of IBQM from the minimum transport intensity is small, ranging from $0.14\%$ to $4.40\%$. With the proposed iterative approach considering different factors, the decomposition strategy does not lead to a substantial loss of solution quality and enables the superior performance of IBQM.

\begin{figure}[pos=t]
\centering
\includegraphics[width=.9\linewidth]{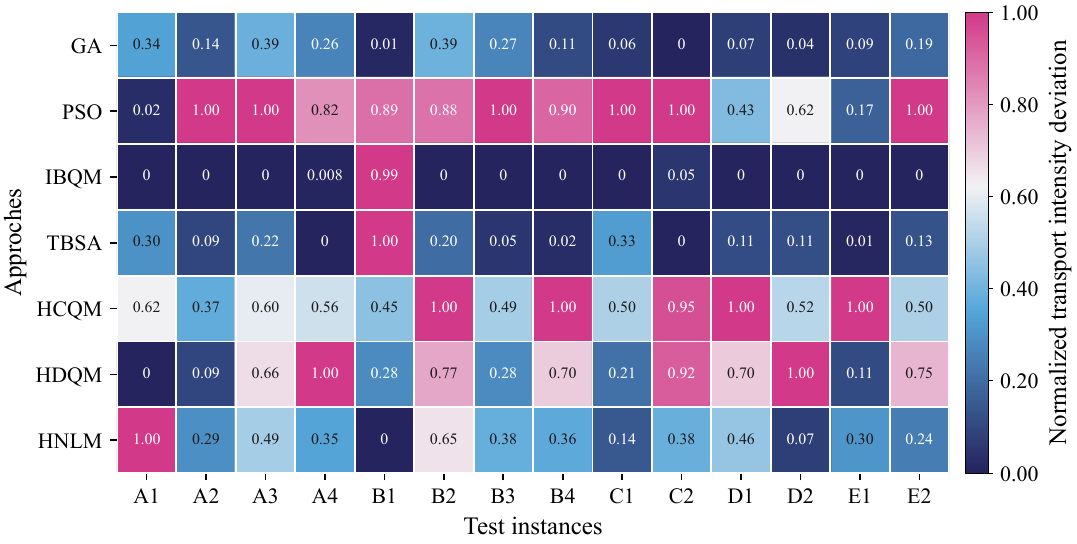}
\caption{Normalized deviation from the minimal transport intensity among all approaches.}
\label{fig:normalized_ti_deviation}
\end{figure}

Fig.~\ref{fig:normalized_ti_deviation} shows the normalized deviation from the minimal transport intensity among all compared approaches. For each instance, the normalized deviation is calculated according to Eq.~\eqref{eq:ti_norm}.
\begin{equation}
\label{eq:ti_norm}
\tilde{TI}=
\dfrac{TI-TI_{\min}}{TI_{\max}-TI_{\min}}
\end{equation}
Here, $TI_{\min}$ and $TI_{\max}$ denote the minimum and maximum transport intensity obtained by all compared approaches for the corresponding instance.

HDQM obtains the same minimum transport intensity as IBQM in A1 and remains competitive in A2, B1, B3, and E1, which are small-sized or medium-sized instances with lower structural complexity. However, HDQM obtains the worst solution quality in several large-scale instances. This is attributable to the candidate selection procedure, which improves scalability but also limits the search space. Although HCQM has a larger search space and better scalability, it does not obtain the minimum transport intensity in any instance and shows higher normalized transport intensity deviation in most cases. Among the QA-based approaches that solve the problem directly, HNLM provides the best solution quality in $71.4\%$ of the instances. Compared with the classical methods, the superiority in solution quality of HNLM is mainly shown in comparison with PSO, while HNLM maintains moderate normalized transport intensity deviation in most instances except A1.

\begin{figure}[pos=t]
\centering
\includegraphics[width=.9\linewidth]{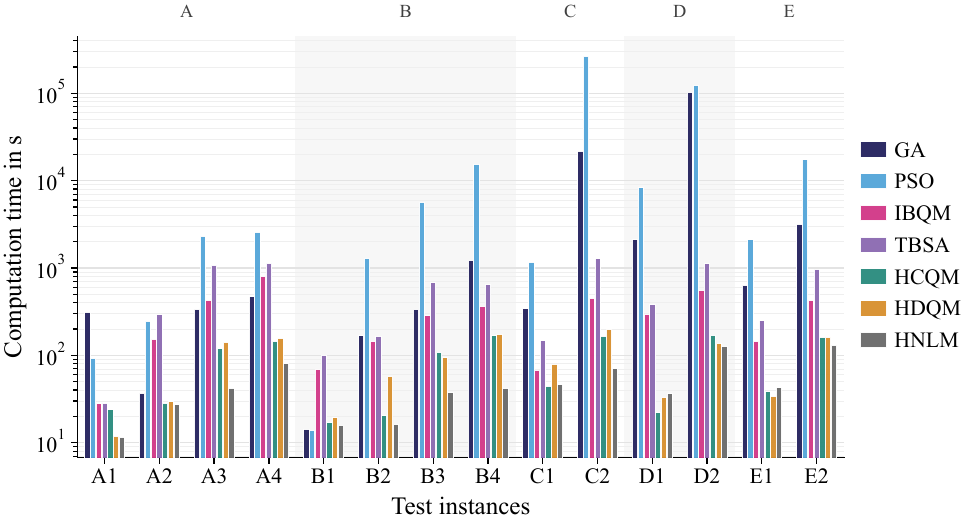}
\caption{Computation times of all approaches.}
\label{fig:compt_time}
\end{figure}

In terms of computation time, the QA-based approaches show a substantial advantage over the classical methods across the evaluated instances. Fig.~\ref{fig:compt_time} illustrates the computation-time comparison of all approaches. Compared with the classical methods, the fastest QA-based approach requires less computation time in $92.9\%$ of the instances. Moreover, the QA-based approaches that solve the problem directly require substantially less time than IBQM and solve all instances in less than three minutes. Among them, HNLM shows the shortest computation time in $78.6\%$ of the instances and a maximum reduction of $94.51\%$ relative to the best-performing classical method.

IBQM requires longer computation times because the BQM formulation contains many quadratic interactions and the iterative approach requires model submission and result transfer in each iteration. However, IBQM obtains a solution within 15 minutes for all instances, which is substantially faster than the classical methods in large and structurally complex instances. In C2, GA obtains the minimum transport intensity but requires almost one day, while PSO requires more than three days. In contrast, IBQM obtains a solution with a normalized transport intensity deviation of only $0.05$ within less than 10 minutes. Compared with TBSA using the same model input, IBQM improves the transport intensity in $85.7\%$ of the instances, with a maximum reduction of $4.51\%$, and reduces computation time in all instances, with a maximum reduction of $64.75\%$. This indicates the superior performance of QA.

In summary, the investigated QA-based approaches can solve factory layout planning problems with up to 75 functional units and different manufacturing-related characteristics within 15 minutes. The comparison further indicates formulation-specific performance differences among the QA-based approaches with respect to solution quality, computation time, and scalability. HCQM and HNLM provide the strongest scalability potential, whereas the scalability of IBQM and HDQM is mainly limited by the large number of quadratic interactions. In terms of solution quality, IBQM provides the strongest performance, both compared with the other QA-based approaches and with the classical reference methods in most instances. The other QA-based approaches do not show superiority in solution quality over the classical methods, but they provide substantial advantages in computation time. Among these approaches, HNLM shows the best overall performance among the QA-based approaches that solve the problem directly, with short computation times and moderate normalized transport intensity deviation in most instances. These results demonstrate the practical applicability of QA-based methods for factory layout planning across problems with varying characteristics.


\section{Conclusion and outlook}
\label{sec:conclusion}

This paper investigates the applicability of various QA-based approaches to early-stage factory layout planning. The problem formulation includes single and composite rectangular functional units, non-overlapping and minimum-distance requirements, infeasible placement areas, and transport route areas, with transport intensity as the optimization objective. Based on a common problem definition, four QA-based approaches are developed and evaluated, namely IBQM, HDQM, HCQM, and HNLM. To improve scalability under the limitations of current QA solvers, an iterative decomposition strategy is developed for IBQM, while a candidate selection procedure is introduced for HDQM. The approaches are evaluated on benchmark instances with increasing scale and structural complexity and are compared with GA, PSO, and TBSA as classical reference methods.

The results demonstrate the superior solution quality and computational efficiency of QA-based approaches over classical metaheuristic methods for factory layout planning, especially in larger and more complex layout instances. Moreover, the comparison shows that different QA-based approaches provide different advantages. IBQM achieves the highest solution quality across most instances through the proposed iterative decomposition strategy, while HCQM and HNLM show stronger scalability potential due to their more compact solver representations. In particular, among the QA-based approaches that solve the complete problem directly, HNLM achieves the best balance of short computation times and stable solution quality. These results indicate that QA-based approaches can support complex factory layout planning by providing solver-specific advantages in solution quality, scalability, and computation time.

However, in the practical manufacturing system, the objective is not limited to the minimization of transport intensity. Future research should extend the current approaches to multi-objective layout planning, considering additional objectives such as area utilization, material handling cost, and energy consumption associated with material handling. Furthermore, the current study addresses a static layout problem, in which material flow, the availability of functional units, infeasible placement areas, and transport routes are fixed during the planning procedure. In practical use cases, these inputs can change over time, which motivates the extension of the QA-based approaches to dynamic factory layout planning to address changing production conditions. In addition, future research should investigate the applicability of QA-based approaches to manufacturing systems with more diverse structural, operational, and material-flow characteristics. These include flexible job-shop manufacturing systems with dynamic transport routes driven by varying job assignments, cellular manufacturing systems in which functional unit grouping introduces additional structural constraints, and circular-economy-oriented systems in which remanufacturing, disassembly, and inspection activities extend the material-flow structure.


\section*{Acknowledgement}
This research was funded by the German Research Foundation
(DFG), grant number AU 185/127-1.
\appendix
\renewcommand{\theequation}{\thesection.\arabic{equation}}
\renewcommand{\thefigure}{\thesection.\arabic{figure}}
\renewcommand{\thetable}{\thesection.\arabic{table}}
\makeatletter
\@addtoreset{equation}{section}
\@addtoreset{figure}{section}
\@addtoreset{table}{section}
\@ifundefined{theHequation}{}{\renewcommand{\theHequation}{appendix.\thesection.\arabic{equation}}}
\@ifundefined{theHfigure}{}{\renewcommand{\theHfigure}{appendix.\thesection.\arabic{figure}}}
\@ifundefined{theHtable}{}{\renewcommand{\theHtable}{appendix.\thesection.\arabic{table}}}
\makeatother
\section{Auxiliary distance terms for HDQM candidate selection}
\label{app:dqm_candidate_selection}

The auxiliary distance terms used in Eq.~\eqref{eq:dqm_case_score_terms} are given as follows.
\begin{equation}
\label{eq:app_dqm_tp_distance}
d^{TP}_{u,i,s}
=
\operatorname{dist}_1
\left(
(x_{u,i}(s),y_{u,i}(s)),TP
\right)
\end{equation}

\begin{equation}
\label{eq:app_dqm_io_distance}
\begin{aligned}
d^{I}_{u,s}
&=
\frac{1}{|R_u|}
\sum_{i\in R_u}
\left(
|x_{u,i}(s)-x_I|
+
|y_{u,i}(s)-y_I|
\right)\\
d^{O}_{u,s}
&=
\frac{1}{|R_u|}
\sum_{i\in R_u}
\left(
|x_{u,i}(s)-x_O|
+
|y_{u,i}(s)-y_O|
\right)
\end{aligned}
\end{equation}

\begin{equation}
\label{eq:app_dqm_uu_distance}
d^{UU}_{u,u',s}
=
\min_{s'\in S_{u'}}
\frac{1}{|R_u||R_{u'}|}
\sum_{i\in R_u}
\sum_{j\in R_{u'}}
\left(
|x_{u,i}(s)-x_{u',j}(s')|
+
|y_{u,i}(s)-y_{u',j}(s')|
\right)
\end{equation}

\section{Supplementary Material}
\label{app:supp_instance_data}
The data of all test instances can be found in \url{https://seafile.rlp.net/d/0a43c44f52104249945c/}.

\printcredits

\bibliographystyle{elsarticle-num}

\bibliography{references}
\end{document}